\documentstyle[11pt]{article}
\def\g{{\,\rm \gamma \,}}
\def\sign{{\,\rm sign \,}}

\def\T{{\cal T}}
\def\Id{{\,\rm Id \,}}

\def\CC{{\,\rm C\,}}
\def\WW{{\,\rm W\,}}
\def\qiq{{\quad\mbox{in}\quad}}

\def\o{{\,\rm o \,}}
\def\id{{\,\rm id \,}}

\def\sym{{\,\rm sym \,}}

\def\ii{{\,\rm i \,}}
\def\dist{{\,\rm dist \,}}

\def\SO{{\,\rm SO \,}}

\def\B{{\,\cal B \,}}

\def\+M{{\,\rm M^{n\times n}_+ \,}}
\def\tr{{\,\rm tr \,}}

\def\qfq{{\quad\mbox{for}\quad}}

\def\ii{{\,\rm i \,}}

\def\C{{\bf C }}

\def\O{{\,\rm O\,}}

\def\lam{\lambda}

\def\E{{\cal E}}

\def\X{{\cal X}}
\def\Q{{\cal Q}}

\def\E{{\cal E}}

\def\V{{\cal V}}
\def\F{{\cal F}}

\def\O{{\cal O}}

\def\H{{\cal H}}

\def\F{{\cal F}}
\def\Q{{\cal Q}}

\def\V{{\cal V}}

\newfont{\Blackboard}{msbm10 scaled 1200}

\newfont{\roma}{cmr10 scaled 1200}

\def\<{{\langle}}
\def\>{{\rangle}}

\def\g{\gamma}
\def\Ga{\Gamma}
\def\var{\varphi}
\def\si{\sigma}

\def\a{\alpha}
\def\b{\beta}

\def\Om{\Omega}

\newtheorem{thm}{{}\hskip\parindent Theorem}[section]
\newtheorem{lem}{{}\hskip\parindent Lemma}[section]
\newtheorem{pro}{{}\hskip\parindent Proposition}[section]

\newtheorem{dfn}{{}\hskip\parindent Definition}[section]
\newtheorem{rem}{{}\hskip\parindent Remark}[section]

\def\dfrac{\displaystyle\frac}

\def\pl{\partial}
\def\rw{\rightarrow}

\def\na{\nabla}

\def\be{\begin{equation}}
\def\ee{\end{equation}}
\def\beq{\arraycolsep=1.5pt\begin{eqnarray}}
\def\eeq{\end{eqnarray}}

\def\Up{\Upsilon}
\def\z{\zeta}

\def\R{I\!\!R}

\def\n{\vec{n}}
\large
\title{Linear Strain Tensors on Hyperbolic Surfaces and Asymptotic Theories for Thin  Shells}
\date{}
\author{
Peng-Fei YAO\\[0.3cm]
Key Laboratory of  Systems and Control\\
Institute of Systems Science,
Academy of Mathematics and Systems Science\\
Chinese Academy of Sciences, Beijing 100190, P. R. China\\
School of Mathematical Sciences\\
University of Chinese Academy of Sciences, Beijing 100049, China\\
e-mail: pfyao@iss.ac.cn}

\begin{document}
\maketitle
 \footnote{This work is  supported by the National
Science Foundation of China, grants  no. 61473126 and no. 61573342, and Key Research Program of Frontier Sciences, CAS, no. QYZDJ-SSW-SYS011.}

\begin{quote}
\begin{small}
{\bf Abstract} \,\,\,We perform a detailed analysis of the solvability of linear strain equations
on hyperbolic surfaces under a technical assumption (noncharacteristic). For regular enough hyperbolic
surfaces, it
is proved that smooth infinitesimal isometries are dense in the $W^{2,2}$ infinitesimal isometries  and that smooth enough infinitesimal isometries can be matched with higher
order infinitesimal isometries. Then those results are applied to elasticity of thin shells for the $\Ga$-limits. The recovery sequences ($\Ga$-lim sup inequlity) are obtained for dimensionally-reduced shell theories, when the elastic energy density scales like $h^\b,$ $\b\in(2, 4),$ that is, intermediate
regime between ¡±pure bending¡± ($\b=2$) and the von-K¡äarm¡äan regime ($\b=4$), where $h$ is thickness of a shell.
\\[3mm]
\end{small}
\end{quote}

\setcounter{equation}{0}
\section{Introduction and main results}
\def\theequation{1.\arabic{equation}}
\hskip\parindent
Let $M\subset\R^3$ be a surface with a normal $\n$ and let the middle surface of a shell be  an open set $\Om\subset M.$ Let $T^k(M)$ denote  all the $k$-order tensor fields on $M$ for an integer $k\geq0.$ Let $T^2_{\sym}(M)$ be all the $2$-order symmetrical tensor fields on $M.$ For $y\in H^1(\Om,\R^3),$ we decompose it into $y=W+w\n,$ where $w=\<y,\n\>$ and $W\in T(\Om).$
For $U\in T^2_\sym(\Om)$ given, linear strain tensor of a displacement $y\in\WW^{1,2}(\Om,\R^3)$ of the middle surface $\Om$ takes the form
\be \sym DW+w\Pi=U\qfq x\in\Om,\label{01}\ee where $D$ is the connection of the induced metric in $M,$  $2\sym DW=DW+D^TW,$ and $\Pi$ is the second fundamental form of $M.$
Equation (\ref{01}) plays a fundamental role in the theory of thin shells, see \cite{HoLePa, LePa,LeMoPa, LeMoPa1} . When $U=0,$ a solution $y$ to (\ref{01}) is referred to as an {\it infinitesimal isometry.} Under a technical assumption (noncharacteristic) on hyperbolic surfaces, we establish existence, uniqueness, and regularity of solutions for (\ref{01}).
For regular enough hyperbolic
surface that satisfies a noncharacteristic assumption, it
is proved that smooth infinitesimal isometries are dense in the $W^{2,2}(\Om,\R^3)$ infinitesimal isometries (Theorem \ref{t1.1}) and that smooth enough infinitesimal isometries can be matched with higher
order infinitesimal isometries (Theorem \ref{t1.2}). This ¡±matching property¡± is an important tool in
obtaining recovery sequences ($\Ga$-lim sup inequlity) for dimensionally-reduced shell theories
in elasticity, when the elastic energy density scales like $h^\b,$ $\b\in(2, 4),$ that is, intermediate
regime between ¡±pure bending¡± ($\b=2$) and the von-K¡äarm¡äan regime ($\b=4$). Such results have been obtained for elliptic surfaces \cite{LeMoPa1} and developable surfaces \cite{HoLePa}.
A survey on this topic is presented in \cite{LePa}. Here we shall establish the similar results for hyperbolic surfaces in
Theorems \ref{t1.5}-\ref{t1.6}.

We state our main results for the hyperbolic surfaces as follows.

A region $\Om\subset M$ is said to be hyperbolic if its Gaussian curvature $\kappa$ is strictly negative. We assume throughout this paper that
$$\kappa(x)<0\qfq x\in\overline{\Om}.$$
 We introduce the notion of a {\it noncharacteristic region} below, subject to the second fundamental form $\Pi$ of the surface $M.$

\begin{dfn}
A region $\Om\subset M$ is said to be
{\it noncharacteristic} if
$$ \Om=\{\,\a(t,s)\,|\,(t,s)\in(0,a)\times(0,b)\,\},$$ where $\a:$ $[0,a]\times[0,b]\rw M$ is an imbedding map which is a family of regular curves with two parameters $t,$ $s$ such that
$$ \Pi(\a_t(t,s),\a_t(t,s))\not=0,\quad\mbox{for all}\quad (t,s)\in[0,a]\times[0,b],$$
$$ \Pi(\a_s(0,s),\a_s(0,s))\not=0,\quad \Pi(\a_s(a,s),\a_s(a,s))\not=0,\quad\mbox{for all}\quad s\in[0,b],$$
$$ \Pi(\a_t(0,s),\a_s(0,s))=\Pi(\a_t(a,s),\a_s(a,s))=0,\quad\mbox{for all}\quad s\in[0,b].$$
\end{dfn}

Consider a surface given by the graph of a function $h: $ $\R^2\rw\R,$
$$M=\{\,(x,h(x))\,|\,x=(x_1,x_2)\in\R^2\,\}.$$ Under the coordinate system $\psi(p)=x$ for $p=(x,h(x))\in M,$
$$\pl x_1=(1,0,h_{x_1}(x)),\quad \pl x_2=(0,1,h_{x_2}(x)),\quad \n=\frac{1}{\sqrt{1+|\nabla h|^2}}(-\nabla h, 1),$$
$$\Pi=-\frac{1}{\sqrt{1+|\nabla h|^2}}\nabla^2h,\quad \kappa=\frac{h_{x_1x_1}h_{x_2x_2}-h_{x_1x_2}^2}{(1+|\nabla h|^2)^2}.$$

(i)\,\,\,Let $h(x)=h_1(x_1)+h_2(x_2)$ where $h_i:$ $\R\rw\R$ are $\CC^2$ functions with $h''_1(x_1)h''_2(x_2)<0.$
Let $\si_i\in\R$ for $1\leq i\leq4$ with $\si_1<\si_2$ and $\si_3<\si_4.$ Then
$$\Om=\{\,(x,h(x))\,|\,\si_1<x_1<\si_2,\,\,\si_3<x_2<\si_4\,\}$$ is noncharacteristic.

(ii)\,\,\,Let $h(x)=x_1^3-3x_1x_2^2.$ Then
$$\kappa(p)<0\qfq p=(x,h(x)),\quad x\in\R^2,\quad |x|>0.$$
For $\varepsilon>0$  and $\si_1<\si_2$ given
$$\Om=\{\,(x,h(x))\,|\,\varepsilon<x_1<\frac{1}{\varepsilon},\,\,\frac{\si_1}{x_1}<x_2<\frac{\si_2}{x_1}\,\}$$ is noncharacteristic. However, there exists a region on $M$ for the $h,$ that is not a  noncharacteristic. For example,  a region
$$\Om=\{\,(x,h(x))\,|\,a<|x|<b\,\}$$ is not  noncharacteristic, where $0<a<b$ and $|x|=x_1^2+x_2^2,$ since its boundaries $|x|=a$ and $|x|=b$ are not  noncharacteristic curves.

Moreover, if $\Om$ is given by a single principal coordinate, i.e.,
$$\Om=\{\,\a(t,s)\,|\,(t,s)\in(a,b)\times(c,d)\,\},$$ where
$$\nabla_{\pl t}\n=\lam_1\pl t,\quad \nabla_{\pl s}\n=\lam_2\pl s,\quad \lam_1>0,\quad\lam_2<0,$$ and $\n$ is the normal of $M,$ then $\Om$ is a noncharacteristic region clearly. Since a principal coordinate exists locally \cite{Yao20181}, a noncharacteristic region also exists locally.

The notion of the noncharacteristic region is  a technical assumption and a different region is given in \cite{Yao2018}. In general, for $U\in T^2_\sym(\Om)$ given, there are many solutions to (\ref{01}). The aim of this assumption is to help us choose a regular solution for each $U.$ In fact, equation (\ref{01}) can be translated into a scalar second order partial differential  equation (see Theorem \ref{t2.1} later), which is elliptic for the elliptic surface \cite{LeMoPa1},  parabolic for the developable surface with no flat part \cite{HoLePa}, and hyperbolic for the hyperbolic surface, respectively.
 Here we assume  $\Om$ to be a  noncharacteristic region in order to set up appropriate boundary conditions such that the scalar equation is to be  well-posedness (see Theorem \ref{t4.1}  and (\ref{3.1})-(\ref{x1})).  The main observation is that if the values of a solution $v$  to the hyperbolic equation (\ref{2.17}) and its derivatives along a noncharacteristic curve are preset, then the solution $v$ is uniquely determined in some neighborhood of this curve. We first solve (\ref{2.17}) locally and then paste up the local solutions to yield a global one (see Lemma \ref{l4.4}), where the noncharacteristic  assumption is such that this produce is successful.   We believe the corresponding results hold true for a more general region but some more complicated discusses may be involved.\\

We say that a noncharacteristic  region $\Om\subset M$ is {\it of class $\CC^{m,1}$} for some integer $m\geq0$ if the surface $M$ is of class $\CC^{m,1}$ and  all the curves $\a(0,\cdot),$ $\a(a,\cdot),$ and $\a(\cdot,s)$  for each $s\in[0,b]$ are  of class $\CC^{m,1}.$  The points $\a(0,0),$ $\a(a,0),$ $\a(0,b),$ and $\a(a,b)$ are angular points of $\Om$ even if $\Om$ is smooth.

\begin{thm}\label{t00}Let $\Om$ be a noncharacterisic  region of class $\CC^{2,1}.$ For $U\in\CC^{1,1}(\Om,T^2_{\sym}),$ there exists a solution $y=W+w\n\in\CC^{0,1}(\Om,\R^3)$ to equation $(\ref{01})$ satisfying the bounds
\be \|W\|_{\CC^{1,1}(\Om,T)}+\|w\|_{\CC^{0,1}(\Om)}\leq C\|U\|_{\CC^{1,1}(\Om,T^2_{\sym})}.\label{t0}\ee
If, in addition, $\Om\in\CC^{m+2,1},$ $U\in\CC^{m+1,1}(\Om,T^2_{\sym})$ for some $m\geq1,$ then
\be \|W\|_{\CC^{m+1,1}(\Om,T)}+\|w\|_{\CC^{m,1}(\Om)}\leq C\|U\|_{\CC^{m+1,1}(\Om,T^2_{\sym})}.\label{t2}\ee
\end{thm}

\begin{rem} For the solvability of $(\ref{01}),$ the  noncharacterisic assumption of $\Om$ can be relaxed. Let $\Om$ be not a noncharacterisic  region but there be  a noncharacterisic  one $\hat\Om$ such that
$\Om\subset\hat\Om.$ Then Theorem $\ref{t00}$ still holds true. In fact, we can extend  $U$ such that $\hat U\in\CC^{m+1,1}(\hat\Om,T^2_{\sym})$ with the estimate
$$\|\hat U\|_{\CC^{m+1,1}(\hat\Om,T^2_{\sym})}\leq C\|U\|_{\CC^{m+1,1}(\Om,T^2_{\sym})}.$$ Then we solve $(\ref{01})$ on $\hat\Om$ to obtain a solution $y$ for which $(\ref{t2})$ still holds.
\end{rem}
For $y\in\WW^{1,2}(\Om,\R^3),$ we denote the left hand side of equation (\ref{01}) by $\sym\nabla y.$
Let
$$\V(\Om,\R^3)=\{\,V\in\WW^{2,2}(\Om,\R^3)\,|\,\sym\nabla V=0\,\}.$$

\begin{thm}\label{t1.1} Let $\Om$ be a noncharacteristic region of class $\CC^{m+2,1}$  for some  integer $m\geq0.$ Then, for every $V\in\V(\Om,\R^3)$
there exists a sequence $\{\,V_k\,\}\subset\V(\Om,\R^3)\cap \CC^{m,1}(\Om,\R^3)$ such that
\be\lim_{k\rw\infty}\|V-V_k\|_{\WW^{2,2}(\Om,\R^3)}=0.\label{sl}\ee
\end{thm}

A one parameter family $\{\,u_\varepsilon\,\}_{\varepsilon>0} \subset\CC^{0,1}(\overline{\Om},\R^2)$ is said to be a (generalized)
$m$th order infinitesimal isometry if the change of metric induced by $u_\varepsilon$ is of order $\varepsilon^{m+1},$ that
$$\|\nabla^Tu_\varepsilon\nabla u_\varepsilon-g\|_{L^\infty(\Om,T^2)}=\O(\varepsilon^{m+1})\quad\mbox{as}\quad \varepsilon\rw0,$$ where $g$ is the induced metric of $M$ from $\R^3,$ see \cite{HoLePa}.
A given $m$th order infinitesimal isometry can be modified by higher order
corrections to yield an infinitesimal isometry of order $m_1>m,$ a property to which we
refer to by {\it matching property of infinitesimal isometries}, \cite{HoLePa,LeMoPa1}. This property plays an important role in the construction of a recover sequence in the $\Ga$-limit for thin shells.

\begin{thm}\label{t1.2}Let $\Om$ be a noncharacteristic region of class $\CC^{2m+1,1}.$ Given $V\in\V(\Om,\R^3)\cap\CC^{2m-1,1}(\Om,\R^3),$
 there exists a family $\{\,w_\varepsilon\,\}_{\varepsilon>0}\subset\CC^{1,1}(\Om,\R^3),$ equibounded in $\CC^{1,1}(\Om,\R^3),$ such
that for all small $\varepsilon>0$  the family:
$$ u_\varepsilon=\id+\varepsilon V +\varepsilon^2w_\varepsilon$$
is a  $m$th order infinitesimal isometry of class $\CC^{1,1}.$
\end{thm}

Let $A:$ $\Om\rw\R^{3\times3}$ be a matrix field. We define $A\in T^2(\Om)$ by
$$A(\a,\b)=\<A(x)\a,\b\>\qfq\a,\,\,\b\in T_x\Om,\quad x\in\Om.$$
For $V\in\V(\Om,\R^3)$ given, there exists a unique $A\in\WW^{1,2}(\Om,T^2)$ such that
\be\nabla_\a V=A(x)\a\qfq \a\in T_xM,\,\,\,\quad A(x)=-A^T(x),\quad x\in\Om.\label{1.5}\ee
The finite strain space is the following closed subspace of $L^2(\Om,T^2_{\sym})$
$$\B(\Om,T^2_{\sym})=\{\,\lim_{h\rw0}\sym\nabla w_h\,|\,w_h\in\WW^{1,2}(\Om,R^3)\,\}$$ where limits are taken in $L^2(\Om,T^2_{\sym}),$ see \cite{GeSa,LeMoPa, Sa}.   $\B(\Om,T^2_{\sym})$ and
$\V(\Om,\R^3)$ are two basic spaces for the $\Ga$-limit functional. A region $\Om\subset M$ is  said to be
{\it approximately robust} if
$$(A^2)_{\tan}\in\B(\Om,T^2_{\sym})\qfq V\in\V(\Om,\R^3),$$ where
$$(A^2)_{\tan}(\a,\b)=\<A^2\a,\b\>\qfq\a,\,\,\b\in T_x\Om,\quad x\in\Om.$$
If $\Om$ is approximately robust, then the $\Ga$-limit functional can be simplified to the bending energy. An  approximately robust surface exhibits a better capacity to resist
 stretching so that the limit functional  consists only of a bending term, see \cite{LeMoPa}.

\begin{thm}\label{t1.3} Let $\Om\subset M$ be a noncharacteristic region of class $\CC^{2,1}.$ Then $\Om$ is approximately robust.
\end{thm}

{\bf Application to elasticity of thin shells}\,\,\,
Let $\n$ be the normal field of surface $M.$ Consider a family $\{\,\Om_h\,\}_{h>0}$ of thin shells of thickness $h$ around $\Om,$
$$\Om_h=\{\,x+t\n(x)\,|\,x\in\Om,\,\,|t|<h/2\,\},\quad 0<h<h_0,$$ where $h_0$ is small enough so that the projection map $\pi:$  $\Om_h\rw\Om,$ $\pi(x+t\n)=x$ is well defined.
For a $\WW^{1,2}$ deformation $u_h:$ $\Om_h\rw\R^3,$ we assume that its elastic energy (scaled per
unit thickness) is given by the nonlinear functional:
$$E_h(u_h)=\frac{1}{h}\int_{\Om_h}W(\nabla u_h)dz.$$
The stored-energy density function $W:$ $\R^3\times\R^3\rw\R$ is $\CC^2$ in an open neighborhood of
SO(3), and it is assumed to satisfy the conditions of normalization, frame indifference and
quadratic growth: For all $F\in\R^3\times\R^3,$ $R\in\SO(3),$
$$W(R)=0,\quad W(RF)=W(F),\quad W(F)\geq C\dist^2(F,\SO(3)),$$
with a uniform constant $C>0.$ The potential $W$ induces the quadratic forms (\cite{FrJaMu})
$$\Q_3(F)=D^2W(Id)(F,F),\quad\Q_2(x,F_{\tan})=\min\{\,Q_3(\hat F)\,|\,\hat F=F_{\tan}\,\}.$$

We shall consider a sequence $e_h>0$ such that:
\be 0<\lim_{h\rw0}e_h/h^\b<\infty\quad\mbox{for some $2<\b\leq4.$}\label{x1.3}\ee
Let
$$\b_m=2+2/m.$$

Recall the following
result.
\begin{thm}\label{t1.4}$\cite{LeMoPa}$\label{t1.4}
Let $\Om$ be a surface embedded in $\R^3,$ which is compact, connected, oriented,
of class $\CC^{1,1},$ and whose boundary $\pl\Om$ is the union of finitely many Lipschitz curves.
Let $u_h\in\WW^{1,2}(\Om_h,\R^3)$ be a sequence of deformations whose scaled energies $E_h(u_h)/e_h$ are
uniformly bounded. Then there exist a sequence $Q_h\in\SO(3)$ and $c_h\in\R^3$ such that for the
normalized rescaled deformations
$$y_h(z)=Q_hu_h(x+\frac{h}{h_0}t\n(x))-c_h,\quad z=x+t\n(x)\in\Om_{h_0},$$ the following holds.

$(i)$ $y_h$ converge to $\pi$ in $\WW^{1,2}(\Om_{h_0},\R^3).$

$(ii)$ The scaled average displacements
$$ V_h(x)=\frac{h}{h_0\sqrt{e_h}}\int_{-h_0/2}^{h_0/2}[y_h(x+t\n)-x]dt$$
converge to some $V\in\V(\Om,\R^3).$

$(iii)$ $\lim\inf_{h\rw0} E_h(u_h)/e_h\geq I(V),$ where
\be I(V)=\frac{1}{24}\int_\Om\Q_2\Big(x,(\nabla(A\n)-A\nabla\n)_{\tan}\Big)dg,\label{1.4}\ee where $A$ is given in $(\ref{1.5}).$
\end{thm}

The above result proves the lower bound for the $\Ga$-convergence. We now state the upper bound in the $\Ga$-convergence result for a smooth noncharacteristic region.

 Since $\Om$ is  approximately robust (Theorem \ref{t1.3}), Theorem \ref{t1.5} below follows from  \cite[Theorem 2.3]{LeMoPa} immediately.
\begin{thm}\label{t1.5}
Let $\Om\subset M$ be a noncharacteristic region of class $\CC^{2,1}.$ Assume that $(\ref{x1.3})$ holds for $\b=4.$
Then for every $V\in\V(\Om,\R^3)$ there exists a sequence of deformations $\{\,u_h\,\}\subset\WW^{1,2}(\Om,\R^3)$ such that $(i)$ and $(ii)$ of Theorem $\ref{t1.4}$ hold. Moreover,
\be \lim_{h\rw0}\frac{1}{e_h}E_h(u_h)=I(V),\label{lim}\ee where $I(V)$ is given in $(\ref{1.4}).$
\end{thm}

\begin{thm} \label{t1.6} Let assumption $(\ref{x1.3})$ hold with $2<\b<4.$ Let $\Om\subset M$ be a noncharacteristic region of class $\CC^{2m+1,1},$ where $m\geq2$ is given such that
$$e_h=\o(h^{\b_m}).$$
Then the results in Theorem $\ref{t1.5}$ hold.
\end{thm}

The rest of the paper is organized as follows.

Section 2 reduces the linear strain equations (\ref{01})
into one scalar  second order equation (\ref{2.17}) (Theorem \ref{t2.1}).

Sections 3  makes preparations to solve problem (\ref{2.17}).  The main observation is that  under an asymptotic coordinate system, this equation locally takes a normal form (Proposition \ref{p4.1}). Thus we studies solvability regions for the normal equation, in where  existence, uniqueness
and estimates for solutions are presented.

Section 4 is devoted to solvability of the scalar equation (\ref{2.17}).  Using solvability of a normal equation in Section 3, we first solve the scalar equation (\ref{2.17}) locally and then path the local solutions together (Theorems
4.1-4.4), where the noncharacteristic assumption of the region $\Om$ is needed to guarantee this process to be successful.

Section 5 returns to the main theorems  in Section 1, and provides proofs for them, using the main results in Section 4.

\setcounter{equation}{0}
\section{ Linear Strain Equations}
\def\theequation{2.\arabic{equation}}
\hskip\parindent We reformulate some expressions from \cite[Section 9.2]{HH} to reduce (\ref{01}) to a coordinate free, scalar equation which can be solved by selecting special charts.

Let $k\geq1$ be an integer. Let $T\in T^k(M)$ be a $k$th order tensor field and let $X\in T(M)$ be a vector field. We define a $k-1$th order tensor field by
$$\ii_XT(X_1,\cdots,X_{k-1})=T(X,X_1,\cdots,X_{k-1})\qfq X_1,\,\,\cdots,\,X_{k-1}\in T(M),$$ which is called an {\it inner product} of $T$ with $X.$ For any $T\in T^2(M)$ and $\a\in T_xM,$
$$\tr_g\ii_\a  DT$$ is a linear functional on $T_xM,$ where $\tr_g\ii_\a DT$ is the trace of the 2-order tensor field $\ii_\a DT$ in the induced metric $g.$ Thus there is a vector, denoted by $\Lambda(T),$ such that
\be\<\Lambda(T),\a\>=\tr_g\ii_\a DT\qfq \a\in T_xM,\,\,x\in M.\label{xx1}\ee Clearly, the above formula defines a vector field $\Lambda(T)\in T(M).$

We also need another linear operator $Q$ as follows. Let $M$ be oriented and $\E$ be the volume element of $M$ with the positive orientation. Let $x\in M$ be given and let $e_1,$ $e_2$ be an orthonormal basis of $T_xM$  with the positive orientation, that is,
$$\E(e_1,e_2)=1\quad \mbox{at}\quad x.$$ We define $Q:$ $T_xM\rw T_xM$ by
\be Q\a=\<\a,e_2\>e_1-\<\a,e_1\>e_2\quad\mbox{for all}\quad\a\in T_xM.\label{1.3n}\ee $Q$ is well defined in the following sense: Let $\hat e_1,$ $\hat e_2$ be a different orthonormal basis of $T_xM$  with the positive orientation,
$$\E(\hat e_1,\hat e_2)=1.\label{1.3}$$ Let
$$\hat e_i=\sum_{j=1}^2\a_{ij}e_j\qfq i=1,\,\,2.$$ Then
$$1=\E(\hat e_1,\hat e_2)=\a_{11}\a_{22}-\a_{12}\a_{21}.$$ Using the above formula, a simple computation yields
$$ \<\a,\hat e_2\>\hat e_1-\<\a,\hat e_1\>\hat e_2=\<\a, e_2\> e_1-\<\a, e_1\> e_2.$$ Clearly, $Q:$ $T_xM\rw T_xM$ is an isometry and
$$Q^T=-Q,\quad Q^2=-\Id.$$

\begin{rem}
$Q:$ $T_xM\rw T_xM$ is the rotation by $\pi/2$ along the clockwise direction.
\end{rem}
 The operator, defined above, defines an operator,
still denoted by $Q$ : $T(M)\rw T(M),$ by
$$ (QX)(x)=QX(x),\quad x\in M,\quad X\in T(M).$$ For each $k\geq2,$ the operator $Q$ further induces an operator, denoted by $Q^*$: $T^k(M)\rw T^k(M)$ by
\be(Q^*T)(X_1,\cdots, X_k)=T(QX_1,\cdots,QX_k),\quad X_1,\,\cdots,\,X_k\in T(M),\quad T\in T^k(M).\label{nn1.3}\ee
Notice that orientability of $M$ is necessary to operators $Q$ or $Q^*.$

Let $x\in\Om$ be given and let $y\in W^{1,2}(\Om,\R^3).$ Set
\be p(y)(x)=\frac{1}{2}[\nabla y(e_2,e_1)-\nabla y(e_1,e_2)]\qfq x\in\Om,\label{2.1}\ee where $\nabla y(\a,\b)=\<\nabla_\b y,\a\>$ for $\a,$ $\b\in T_xM,$   $\nabla$ is the differential in the Euclidean space $\R^3,$ and $e_1,$ $e_2$ is an orthonormal basis of $T_xM$ with the positive orientation. It is easy to check that the value of the right hand side of (\ref{2.1}) is independent of choice of a positively orientated orthonormal basis. Thus
$$p:\quad W^{1,2}(\Om,\R^3)\rw L^2(\Om)$$ is a linear operator.

For $U\in T^2_{\sym}(M)$ given, consider problem
\be \sym\nabla y (\a,\b)=U(\a,\b)\qfq\a,\quad\b\in T_xM,\quad x\in M,\label{2.2}\ee where $y\in W^{1,2}(\Om,\R^3).$

Let $x\in\Om$ be given. To simplify computation we use many times the following special frame field: Let  $E_1,$ $E_2$ be a positively orientated frame field normal at $x$ with following properties
$$\<E_i,E_j\>=\delta_{ij} \quad\mbox{in some neighbourhood of $x,$} $$
\be D_{E_i}E_j=0,\quad \nabla_{E_i}\n=\lam_iE_i\quad\mbox{at $x$ for $1\leq i,\,j\leq2,$}\label{2.3}\ee  where $\nabla$ is the connection of the Euclidean space $\R^3,$ $D$ is the connection of $M$ in the induced metric, $\n$ is the normal field of $M,$ and $\lam_1\lam_2=\kappa$ is the Gaussian curvature. It follows from (\ref{2.3}) that
\be\Pi(E_i,E_j)=\lam_i\delta_{ij},\quad \nabla_{E_i}E_j=-\lam_i\delta_{ij}\n\quad\mbox{at $x$ for $1\leq i,\,j\leq2,$}\label{2.4}\ee where $\Pi(\a,\b)=\<\nabla_\a\n,\b\>$ is the second fundamental form of $M.$ We need to deal with the relation between the connections $\nabla$ and $D,$ carefully.

 Let $y\in W^{1,2}(\Om,\R^3)$ be a solution
to problem (\ref{2.2}). Then (\ref{2.2}) reads
\be \left\{\begin{array}{l} \nabla y(E_1,E_1)=U(E_1,E_1),\\
\nabla y(E_2,E_1)+\nabla y(E_1,E_2)=2U(E_1,E_2),\\
\nabla y(E_2,E_2)=U(E_2,E_2),\end{array}\right.\quad\mbox{in some neighbourhood of $x.$}\label{2.5}\ee Let
$$v=p(y)$$
and define
\be u=\nabla y(\n,E_1)E_1+\nabla y(\n,E_2)E_2.\label{n2.6}\ee We can check easily that $u$ is a globally defined vector field on $\Om.$ Moreover,
$v$ satisfies
\be v+U(E_2,E_1)=\nabla y(E_2,E_1),\quad v-U(E_1,E_2)=-\nabla y(E_1,E_2) \label{2.6}\ee in some neighbourhood of $x.$ Therefore, $\{v,u\}$ determines $\nabla_\a y$ for $\a\in T_xM,$ that is,
\be\left\{\begin{array}{l}\nabla_{E_1}y=U(E_1E_1)E_1+[v+U(E_1,E_2)]E_2+\<u,E_1\>\n,\\
\nabla_{E_2}y=[-v+U(E_1,E_2)]E_1+U(E_2,E_2)E_2+\<u,E_2\>\n.\end{array}\right.\label{2.7}\ee The relation (\ref{2.7}) can be rewritten  as in a form of coordinate free
$$ \nabla_\a y=\ii_\a U-vQ\a+\<u,\a\>\n\qfq\a\in T_xM,\,\,x\in\Om.$$
The function $v$ and the vector field $u$ are the new dependent variables and we proceed to find the differential equations they satisfy.

Differentiating the first equation in (\ref{2.6}) with respect to $E_2$ and using the relations (\ref{2.3}) and (\ref{2.4}), we have
\beq E_2(v)+DU(E_2,E_1,E_2)&&=\nabla^2y(E_2,E_1,E_2)+\nabla y(\nabla_{E_2}E_2,E_1)\nonumber\\
&&=E_1[\nabla y(E_2,E_2)]-\lam_2\nabla y(\n,E_1)\nonumber\\
&&=DU(E_2,E_2,E_1)-\lam_2\<u,E_1\>\quad\mbox{ at $x,$}\label{2.8}\eeq where the following formula has been used
$$\nabla^2y(E_2,E_1,E_2)=\nabla^2y(E_2,E_2,E_1)\quad\mbox{at $x.$}$$
Similarly, we obtain
\be E_1(v)-DU(E_1,E_2,E_1)=-DU(E_1,E_1,E_2)+\lam_1\<u,E_2\>\quad\mbox{at $x.$}\label{2.9}\ee
Combining (\ref{2.8}), (\ref{2.9}) and (\ref{2.3}) yields
\beq Dv&&=[DU(E_1,E_2,E_1)-DU(E_1,E_1,E_2)]E_1+\lam_1\<u,E_2\>E_1\nonumber\\
&&\quad+[DU(E_2,E_2,E_1)-DU(E_2,E_1,E_2)]E_2-\lam_2\<u,E_1\>E_2\nonumber\\
&&=Q\{[DU(E_2,E_1,E_1)+DU(E_2,E_2,E_2)]E_2-[DU(E_2,E_2,E_2)+DU(E_1,E_1,E_2)]E_2\nonumber\\
&&\quad+[DU(E_1,E_2,E_2)+DU(E_1,E_1,E_1)]E_1-[DU(E_1,E_1,E_1)+DU(E_2,E_2,E_1)]E_1\}\nonumber\\
&&\quad+\nabla\n Qu\nonumber\\
&&=Q[\Lambda(U)-D(\tr_g U)]+\nabla\n Qu\qfq x\in\Om,\label{2.10}\eeq where the operator $Q:$ $T_xM\rw T_xM$ is defined in (\ref{1.3n}), $\Lambda(U)\in\X(\Om)$ is given in (\ref{xx1}), and
$\nabla\n:$ $T_xM\rw T_xM$ is the shape operator, defined by
$$\nabla\n\a=\nabla_\a\n\qfq\a\in T_xM,\,\,x\in M.$$

 Now we proceed to derive
the differential equations for which the function v satisfies. Since
$$\kappa=\Pi(E_1,E_1)\Pi(E_2,E_2)-\Pi^2(E_1,E_2)\quad\mbox{in a neighbourhood of $x,$}$$ from (\ref{2.3}) and (\ref{2.4}) we compute
\beq D\kappa&&=[D\Pi(E_1,E_1,E_1)\lam_2+\lam_1D\Pi(E_2,E_2,E_1)]E_1\nonumber\\
&&\quad+[D\Pi(E_1,E_1,E_2)\lam_2+\lam_1D\Pi(E_2,E_2,E_2)]E_2\quad\mbox{at $x.$}\label{2.12n}\eeq
Using (\ref{2.10}), (\ref{2.3}) and (\ref{2.4}), we have
\beq &&D(\nabla\n Qu)(E_1,E_1)=E_1\<\nabla\n Qu,E_1\>=E_1\<u,Q^T\nabla_{E_1}\n\>\nonumber\\
&&=Du(Q^T\nabla_{E_1}\n,E_1)+\<u,D_{E_1}(Q^T\nabla_{E_1}\n)\>\nonumber\\
&&=\lam_1Du(E_2,E_1)+D\Pi(E_1,E_1,E_1)\<u,E_2\>-D\Pi(E_1,E_1,E_2)\<u,E_1\>\quad\mbox{at $x,$}\label{2.12}\eeq where the symmetry of $D\Pi$ is used.
A similar computation yields
\beq &&D(\nabla\n Qu)(E_2,E_2)=-\lam_2Du(E_1,E_2)\nonumber\\
&&\quad+D\Pi(E_1,E_2,E_2)\<u,E_2\>-D\Pi(E_2,E_2,E_2)\<u,E_1\>\quad\mbox{at $x.$}\label{2.13}\eeq
Multiplying (\ref{2.12}) by $\lam_2$ and (\ref{2.13}) by $\lam_1,$ respectively,  summing them, and using (\ref{2.12n}), we obtain
\be\<D(\nabla\n Qu),Q^*\Pi\>=\kappa[Du(E_2,E_1)-Du(E_1,E_2)]+\<Qu, D\kappa\>.\label{x2.19}\ee
Note that the function $Du(E_2,E_1)-Du(E_1,E_2)$ is globally defined on $\Om$ which is independent of choice of a positively orientated orthonormal basis when
the vector field $u$ is given. From (\ref{2.10}) and (\ref{x2.19}), we obtain
$$\<D^2v,Q^*\Pi\>=\<D\{Q[\Lambda(U)-D(\tr_g U)]\},Q^*\Pi\>+\kappa[Du(E_2,E_1)-Du(E_1,E_2)]+\<Qu, D\kappa\>.$$

Next, let us consider the compatibility conditions which insure that a $y$ to satisfy (\ref{2.7}) exists when the function $v$ and the vector field $u$ are given to satisfy (\ref{2.10}).
We define $B:$ $T_xM\rw T_xM$ for $x\in\Om$ by
\be B\a=\ii_\a U-vQ\a+\<u,\a\>\n\qfq\a\in T_xM.\label{xx2.20}\ee It is easy to check that there is a $y:$ $\Om\rw\R^3$ such that
$$\nabla_\a y=B\a\qfq\a\in T_xM,\,\,x\in\Om$$ if and only if the operator $B$ satisfies
\be\nabla_X(BY)=\nabla_Y(BX)+B[X,Y]\qfq X,\,\,Y\in\X(\Om).\label{xx2.21}\ee
Using (\ref{2.3}), (\ref{2.4}), (\ref{2.9}), and (\ref{xx2.20}), we have
\beq\nabla_{E_1}(BE_2)&&=[DU(E_2,E_1,E_1)-E_1(v)+\lam_1\<u,E_2\>]E_1+DU(E_2,E_2,E_1)E_2\nonumber\\
&&\quad+[Du(E_2,E_1)-\lam_1U(E_2,E_1)+v\lam_1]\n\nonumber\\
&&=DU(E_1,E_1,E_2)E_1+DU(E_2,E_2,E_1)E_2\nonumber\\
&&\quad+[Du(E_2,E_1)-\lam_1U(E_2,E_1)+v\lam_1]\n\quad\mbox{at}\quad x.\label{xx2.22}\eeq Similarly, we obtain
\beq\nabla_{E_2}(BE_1)&&=DU(E_1,E_1,E_2)E_1+DU(E_2,E_2,E_1)E_2\nonumber\\
&&\quad+[Du(E_1,E_2)-\lam_2U(E_1,E_2)-v\lam_2]\n\quad\mbox{at}\quad x.\label{xx2.23}\eeq
It follows from (\ref{xx2.22}) and (\ref{xx2.23}) that the relation (\ref{xx2.21}) holds if and only if
$$  Du(E_2,E_1)-Du(E_1,E_2)+\tr_g U(Q\nabla\n\cdot,\cdot)+v\tr_g\Pi=0\qfq x\in\Om.   $$

Moreover, we assume that
\be\kappa(x)\not=0\quad\mbox{for all}\quad x\in\overline{\Om}.\label{2.22n}\ee
From (\ref{2.10}), we obtain
\be u=Q(\nabla\n)^{-1}Q[\Lambda(U)-D(\tr_g U)]-Q(\nabla\n)^{-1}Dv\qfq x\in\Om.\label{2.11}\ee

The above derivation yields the following.

\begin{thm}$(\cite{HH})$\label{t2.1}\,\,\,Suppose that $(\ref{2.22n})$ holds. Let $v$ be a solution to  problem
\be\<D^2v,Q^*\Pi\>=P(U)-v\kappa\tr_g\Pi+X(v)\qfq x\in\Om,\label{2.17}\ee where
\beq P(U)&&=\<D\{Q[\Lambda(U)-D(\tr_g U)]\},Q^*\Pi\>-\<Q[\Lambda(U)-D(\tr_g U)],(\nabla\n)^{-1}D\kappa\>\nonumber\\
&&\quad-\kappa\tr_g U(Q\nabla\n\cdot,\cdot),\label{2.19}\eeq
\be X=(\nabla\n)^{-1}D\kappa.\label{2.20}\ee
Let $u$ be given by $(\ref{2.11}).$ Then there is a $y$ to satisfy $(\ref{2.2})$ such that $(\ref{2.7})$ holds. Moreover,
$$|\nabla y|^2(x)=|U|^2(x)+2v^2(x)+|u(x)|^2\qfq x\in\Om.$$ If, in addition, $y=W+w\n,$ $w=\<y,\n\>,$ then
$$ u=Dw-\ii_W\Pi,$$
$$ Dw=\ii_W\Pi-Q(\nabla\n)^{-1}Dv+Q(\nabla\n)^{-1}Q[\Lambda(U)-D(\tr_g U)].$$
\end{thm}

\begin{rem}
A solution $y,$ modulo a constant vector, in Theorem $\ref{t2.1}$ is unique when a solution $v$ to $(\ref{2.17})$ is given.
\end{rem}

\begin{rem} If $\Om$ is elliptic and $\Pi>0,$ then $\hat g=\Pi$ is another metric on $\Om.$ From $\cite{Yao2012}$ we have
$$\<D^2v,Q^*\Pi\>=\kappa\Delta_{\hat g}v+\frac{1}{2\kappa}\Pi(QD\kappa,QDv)\qfq x\in\Om,$$ where $\Delta_{\hat g}$ is the Laplacion of the metric $\hat g.$ Thus, in this case equation
$(\ref{2.17})$ becomes
$$ \Delta_gv=\frac{1}{\kappa}P(U)-v\tr_g\Pi+\frac{1}{2\kappa}X(v)\qfq x\in\Om. $$
\end{rem}

\setcounter{equation}{0}
\section{Solvability Regions for Normal Equations}
\def\theequation{3.\arabic{equation}}
\hskip\parindent Under an asymptotic coordinate system, equation (\ref{2.17}) on a hyperbolic surface takes
the form of a normal equation in $\R^2$ locally, such as in (\ref{3.7}) below. Thus  the local solvability of equation (\ref{2.17})  transfers to that of equation (\ref{3.7}) in the Euclidean space
$\R^2.$ We study the solvability of the normal equation (\ref{3.7}) in the space $\R^2$ in this section.

We consider the following normal equation
\be w_{x_1x_2}(x)=\eta(f,  w)\qfq x=(x_1,x_2)\in\R^2\label{3.7}\ee
where
$$\eta(f, w)=f+f_0(x)w(x)+X(w),$$  $f_0$ is a function, and $X=(X_1,X_2)$ is a vector field on $\R^2.$

{\bf Regions $E(\g),$ $R(z,a,b),$ $P_i(\b),$ $\Xi_i(\b,\g)$ and $\Phi(\b,\g,\hat\b)$}\,\,\,In appropriate asymptotic coordinate systems, the problem (\ref{2.17}) can transfer to solvability of (\ref{3.7}) on $E(\g),$ $R(z,a,b),$ $P_i(\b),$ $\Xi_i(\b,\g)$ or $\Phi(\b,\g,\hat\b)$ with appropriate boundary data. We now introduce those regions to establish the corresponding solvability.

\subsection{Regions $E(\g)$ and $R(z,a,b)$}
\hskip\parindent
Let $k\geq0$ be an integer. Let $f_0$ and $X$ be of class $\CC^{k-1,1},$ where $\CC^{-1,1}=L^\infty.$
A curve $\g(t)=(\g_1(t),\g_2(t)):$ $[a,b]\rw\R^2$ is said to be {\it noncharacteristic} if
$$\g'_1(t)\g'_2(t)\not=0\qfq t\in[a,b].$$ We define a linear operator $\F:$ $\R^2\rw\R^2$ by
\be\F x=(x_2,-x_1)\qfq x=(x_1,x_2)\in\R^2.\label{f3.2}\ee

Let $\g(t)=(\g_1(t),\g_2(t)):$ $[0,t_0]\rw\R^2$ be a noncharacteristic curve with $\g'_1(0)\g'_2(0)<0.$ We assume that
\be\g'_1(t)>0,\quad\g'_2(t)<0\qfq t\in[0,t_0].\label{non3.2}\ee Otherwise, we consider the curve $z(t)=\g(-t+t_0)$. Set
\be E(\g) =\{\,(x_1,x_2)\in\R^2\,|\,\g_1\circ\g_2^{-1}(x_2)<x_1<\g_1(t_0),\,\g_2(t_0)<x_2<\g_2(0)\,\}.\label{T(Z,a)}\ee
Consider the boundary data
\be w\circ\gamma(t)=q_0(t),\quad\<\nabla w,\F\dot\g\>\circ\gamma(t)=q_1(t)\qfq t\in(0,t_0).\label{3.8}\ee

Next, we consider a rectangle. For $z=(z_1,z_2)\in\R^2,$ $a>0,$ and $b>0$ given,  let
\be R(z,a,b)=(z_1,z_1+a)\times(z_2,z_2+b).\label{R}\ee
Consider the boundary data
\be w(x_1,z_2)=p_1(x_1),\quad w(z_1, x_2)=p_2(x_2)\label{xnn3.9}\ee for $x_1\in[z_1,z_1+a]$ and $x_2\in[z_2,z_2+b],$ respectively.

Let $f$ be a function with its domain $E.$ For simplicity, we denote  $\|f\|_{\CC^{k,1}}=\|f\|_{\CC^{k,1}(\overline{E})},$ $\|f\|_{\WW^{k,2}}=\|f\|_{\WW^{k,2}(E)},$ and so on.

\begin{pro}\label{p3.1}
Let $q_0$ be of $\CC^{k,1}$ and  $q_1,$ $f$ be of $\CC^{k-1,1},$ respectively. Then problem $(\ref{3.7})$ admits a unique solution  $w\in \CC^{k,1}(\overline{E(\g)})$ with the data $(\ref{3.8}).$
Moreover, there is a $C>0,$ independent of solutions $w,$ such that
\be \|w\|_{\CC^{k,1}}\leq C(\|q_0\|_{\CC^{k,1}}+\|q_1\|_{\CC^{k-1,1}}+\|f\|_{\CC^{k-1,1}}).\label{xnn3.3}\ee
\end{pro}

\begin{pro}\label{p03.4}
Let $p_1$ and $p_2$ be of class $\CC^{k,1}$ with $p_1(z_1)=p_2(z_2).$  Let $f$ be of class $\CC^{k-1,1}.$ Then there is a unique solution $w\in \CC^{k,1}(\overline{R(z,a,b)})$ to $(\ref{3.7})$
with the data $(\ref{xnn3.9})$ satisfying
$$\|w\|_{\CC^{k,1}}\leq C(\|p_1\|_{\CC^{k,1}}+\|p_2\|_{\CC^{k,1}}+\|f\|_{\CC^{k-1,1}}).$$
\end{pro}

The proofs of Propositions \ref{p3.1} and \ref{p3.4} will be given after Lemma \ref{p3.4}.

\begin{lem}\label{l3.1} Let  $T>0$ be given.
There is a $\varepsilon_T>0$ such that if $|\g(0)|\leq T$ and $\max\{\g_1(t_0)-\g_1(0),\g_2(0)-\g_2(t_0)\}<\varepsilon_T,$ Theorem $\ref{p3.1}$ holds true.
\end{lem}

{\bf Proof.}\,\,\,  The proof  is broken into several steps as follows.

{\bf Step 1.}\,\,\,Let $k=0$ and let $w\in C^{0,1}(\overline{E(\g)})$ be a solution to (\ref{3.7}) with the data (\ref{3.8}). It follows from (\ref{3.8})
that
$$ w_{x_1}\circ\gamma(t)=\frac{1}{|\g'(t)|^2}[\g_1'(t)q_0'(t)+\g_2'(t)q_1(t)],\quad w_{x_2}\circ\gamma(t)=\frac{1}{|\g'(t)|^2}[\g_2'(t)q_0'(t)-\g_1'(t)q_1(t)].$$

Let $x=(x_1,x_2)\in E(\g)$ be given. We integrate
(\ref{3.7}) with respect to the first variable $\zeta_1$ over $(\g_1\circ\g_2^{-1}(\zeta_2),x_1)$ for $\zeta_2\in(\g_2\circ\g_1^{-1}(x_1),x_2)$ to have
\beq w_{x_2}(x_1,\zeta_2)&&=w_{x_2}\circ\g(\g_2^{-1}(\zeta_2))+\int_{\g_1\circ\g_2^{-1}(\zeta_2)}^{x_1}\eta(f, w)(\zeta_1,\zeta_2)d\zeta_1.\quad\quad\quad\label{nx3.13}\eeq
Then integrating the above identity over $(\g_2\circ\g_1^{-1}(x_1),x_2)$ with respect to the second variable $\zeta_2$ yields
$$ w(x_1,x_2)=\B(q_0,q_1)+\int_{E(x)}\eta(f,w)d\zeta,$$ where
\be\B(q_0,q_1)=q_0\circ\g_1^{-1}(x_1)+\int_{\g_1^{-1}(x_1)}^{\g_2^{-1}(x_2)}\frac{\g_2'(t)}{|\g'(t)|^2}[\g_2'(t)q_0'(t)-\g_1'(t)q_1(t)]dt,\label{fix3.33}\ee
\be E(x)=\{\,(\zeta_1,\zeta_2)\,|\,\g_1\circ\g_2(\zeta_2)<\zeta_1<x_1,\,\g_2\circ\g_1^{-1}(x_1)<\zeta_2<x_2\,\}.\label{fix3.34}\ee

{\bf Step 2.}\,\,\,We define an operator $I:$ $\CC^{0,1}(\overline{E(\g)})\rw \CC^{0,1}(\overline{E(\g)})$ by
\be I(w)=\B(q_0,q_1)+\int_{E(x)}\eta(f,w)d\zeta\qfq w\in \CC^{0,1}(\overline{E(\g)}).\label{xxnn3.28}\ee
 It is easy to check that
$w\in \CC^{0,1}(\overline{E(\g)})$ solves (\ref{3.7}) with the data (\ref{3.8}) if and only if $I(w)=w.$

Next, we show that there is a $0<\varepsilon_T\leq1$ such that when $|\g(0)|\leq T$ and $0<\max\{\g_1(t_0)-\g_1(0),\g_2(0)-\g_2(t_0)\}<\varepsilon_T,$ the map $I:$ $\CC^{0,1}(\overline{E(\g)})\rw \CC^{0,1}(\overline{E(\g)})$
is contractible. Thus the existence and uniqueness of solutions  in the case $k=0$ follows from Banach's fixed point theorem.

A simple computation shows that for $w\in \CC^{0,1}(\overline{E(\g)})$
\beq [I(w)]_{x_1}&&=\frac{1}{|\g'(t)|^2}[\g_1'(t)q_0'(t)+\g_2'(t)q_1(t)]\Big|_{t=\g_1^{-1}(x_1)}+\int_{\g_2\circ \g_1^{-1}(x_1)}^{x_2}\eta(f,w)(x_1,\zeta_2)d\zeta_2,\nonumber\eeq
\beq[I(w)]_{x_2}&&=\frac{1}{|\g'(t)|^2}[\g_2'(t)q_0'(t)-\g_1'(t)q_1(t)]\Big|_{t=\g_2^{-1}(x_2)}+\int^{x_1}_{\g_1\circ\g_2^{-1}(x_2)}\eta(f, w)(\zeta_1,x_2)d\zeta_1.\nonumber\eeq
The above formulas yield for $w_1,$ $w_2\in \CC^{0,1}(\overline{E(\g)}),$
$$\|I(w_1)-I(w_2)\|_{\CC^{0,1}(\overline{E(\g)})}\leq C_T\max\{\lam,\lam^2\}\|w_1-w_2\|_{\CC^{0,1}(\overline{E(\g)})},$$
 where
$$\lam=\max\{\g_1(t_0)-\g_1(0),\g_2(t_0)-\g_2(0)\},\quad C_T=\|f_0\|_{L^\infty(|x|\leq2T)}+\|X\|_{L^\infty(|x|\leq2T)}.$$ Thus, the map $I:$ $\CC^{0,1}(\overline{E(\g)})\rw C^{0,1}(\overline{E(\g)})$
is contractible if $\lam>0$ is small.

{\bf Step 3.}\,\,\,Consider the case $k=1.$ Let $q_0\in \CC^{1,1}[0,t_0],$ $q_1\in \CC^{0,1}[0,t_0],$ and $f\in \CC^{0,1}(\overline{E(\g))}$ be given. By Step 2, there is a $\varepsilon_T>0$ such that
if $|\g(0)|\leq T$  and $0<\lam<\varepsilon_T,$  problem $(\ref{3.7})$ has a unique solution
$w\in \CC^{0,1}(\overline{E(\g)})$ with the data (\ref{3.8}). A formal computation shows that $u=w_{x_1}$ solves problem
\be u_{x_1x_2}=\eta(\hat f, u)\qfq x\in E(\g),\label{x5}\ee with the data
\be u\circ\gamma(t)=\hat q_0(t),\quad \<\na u,\F\dot\g\>\circ\gamma(t)=\hat q_1(t)\qfq t\in(0,t_0),\label{x6}\ee where
$$\hat f=f_{x_1}+f_{0x_1}w+\nabla_{\pl x_1}X(w),\quad \hat q_0(t)=\frac{1}{|\g'(t)|^2}[\g_2'(t)q_0'(t)+\g_1'(t)q_1(t)],$$
$$\hat q_1(t)=\frac{\g_2'(t)}{\g_1'(t)}\hat q_0'(t)-\frac{|\g'(t)|^2}{\g_1'(t)}\eta(f,w)\circ\g(t).$$
We apply Step 2 to problem (\ref{x5}) and (\ref{x6}) to obtain  $u=w_{x_1}\in \CC^{0,1}(\overline{E(\g)})$ when $0<\lam<\varepsilon_T.$
A similar argument yields $w_{x_2}\in \CC^{0,1}(\overline{E(\g)}).$  Thus $w\in\CC^{1,1}(\overline{E(\g)}).$

By repeating the above procedure, the existence and uniqueness of the solutions in the cases $k\geq2$ are obtained.

{\bf Step 4.}\,\,\,
 Let  map $I:$ $\CC^{k,1}(\overline{E(\g)})\rw \CC^{k,1}(\overline{E(\g)})$ be defined in Step 2 and let $w\in\CC^{k,1}(\overline{E(\g)})$ be the solution to problem $(\ref{3.7})$ with the data
 (\ref{3.8}). Then
 \beq\|w\|_{\CC^{k,1}}&&=\|I(w)\|_{\CC^{k,1}}\leq \|I(0)\|_{\CC^{k,1}}+\|I(w)-I(0)\|_{\CC^{k,1}}\nonumber\\
 &&\leq C(\|q_0\|_{\CC^{k,1}}+\|q_1\|_{\CC^{k-1,1}}+\|f\|_{\CC^{k-1,1}})+C_T\max\{\lam,\lam^2\}\|w\|_{\CC^{k,1}}.\nonumber\eeq Thus, the estimate (\ref{xnn3.3}) follows if $\lam>0$ is small.  \hfill$\Box$

By a similar argument as for Lemma \ref{l3.1}, we have the following lemmas.

\begin{lem}\label{p3.4}
 Let $T>0.$ There is $\varepsilon_T>0$ such that if $|z|\leq T$ and $0<\max\{a,b\}<\varepsilon_T,$ then Proposition $\ref{p3.4}$ holds.
\end{lem}

{\bf Proof of Proposition \ref{p3.1}.}\,\,\,\,We shall show that the assumptions£¬ $|\g(0)|\leq T$ and $\max\{\g_1(t_0)-\g_1(0),\g_2(t_0)-\g_2(0)\}<\varepsilon_T$ in Lemma \ref{l3.1} are unnecessary. Let $T>0$ be given such that
$$E(\g)\subset\{\,x\in\R^2\,|\,|x|\leq T\,\}.$$ Let $\varepsilon_T>0$ be given such that Lemmas \ref{l3.1} and \ref{p3.4} hold. We divide the curve $\g$ into
$m$ parts with the points $\tau_0=0,$ $\tau_0<\tau_1<\cdots<\tau_m=t_0$ such that
$$\quad |\g(\tau_{i+1})-\g(\tau_i)|=\frac{\varepsilon_T}{2},\quad 0\leq i\leq m-2,\quad |\g(t_0)-\g(\tau_{m-1})|\leq\frac{\varepsilon_T}{2}.$$
For simplicity, we assume that
$m=3.$ The other cases can be treated by a similar argument.

In the case of $m=3,$ we have
\beq \overline{E(\g)}&&=(\cup_{i=0}^2\overline{E}_i)\cup(\cup_{i=1}^3\overline{R}_i)\eeq where
$$E_i=\{\,x\in E(\g)\,|\,\g_1(\tau_i)\leq x_1\leq\g_1(\tau_{i+1}),\,\g_2(\tau_{i+1})\leq x_2\leq\g_2(\tau_i)\,\}\quad i=0,\,1,\,2,$$
$$R_1=[\g_1(\tau_1),\g_1(\tau_2)]\times[\g_2(\tau_1),\g_2(0)],\quad R_2=[\g_1(\tau_2),\g_1(t_0)]\times[\g_2(\tau_2),\g_2(\tau_1)],$$
$$R_3=[\g_1(\tau_2),\g_1(t_0)]\times[\g_2(\tau_1),\g_2(0)].$$

From Lemma \ref{l3.1}, problem (\ref{3.7}) admits a unique solution $w_i\in \CC^{k,1}(\overline{E}_i)$ for each $i=0,$ $1,$ and $2,$ respectively,
with the corresponding data and the corresponding estimates. We define $w\in \CC^{k,1}(\cup_{i=0}^2\overline{E}_i)$ by
$$w(x)=w_i(x)\qfq x\in \overline{E}_i\qfq i=0,\,1,\,2.$$ We extend the domain of $w$ from $\cup_{i=0}^3\overline{E}_i$ to $\overline{E(\g)}$ by the following way.
By Lemma \ref{p3.4}, we define $w\in \CC^{k,1}(\overline{R}_{i})$
to be the solution $u_i\in\CC^{k,1}(\overline{R}_{i})$ to problem (\ref{3.7}) with the data
$$u_i(\g_1(\tau_i),x_2)=w_{i-1}(\g_1(\tau_i),x_2)\qfq x_2\in[\g_2(\tau_{i}),\g_2(\tau_{i-1})],$$
$$u_i(x_1,\g_2(\tau_i))=w_{i}(x_1,\g_2(\tau_i))\qfq x_1\in[\g_1(\tau_i),\g_1(\tau_{i+1})], $$ for $i=1,$ and $2,$ respectively. Then we extend $w$ on $\CC^{k,1}(\overline{R}_{3})$ to be
the solution $u_3$ of (\ref{3.7}) with
the data
$$u_3(\g_1(\tau_2),x_2)=u_1(\g_1(\tau_2),x_2)\qfq x_2\in[\g_2(\tau_1),\g_2(0)],$$
$$u_3(x_1,\g_2(\tau_2))=u_2(x_1,\g_2(\tau_2))\qfq x_1\in[\g_1(\tau_2),\g_1(t_0)].$$

To complete the proof, we have to show  that $w$ is a $\CC^{k,1}$ solution on all the connection segments between any two subregions above.  Consider the subregion
$$\overline{\tilde{E}}=\overline{E}_0\cup\overline{E}_1\cup\overline{R}_{1}.$$
Since $|\g(\tau_2)-\g(0)|\leq\varepsilon_T,$ Lemma \ref{l3.1} insures that problem (\ref{3.7}) admits a unique solution $\tilde w\in \CC^{k,1}(\overline{\tilde{E}})$ with the corresponding data. Then the uniqueness implies
that $w(x)=\tilde w(x)$ for $x\in\overline{\tilde E}.$ In particular, $w$ is $\CC^{k,1}$ on the segments $\{\,(\g_1(\tau_1),x_2)\,|\,x_2\in[\g_2(\tau_1),\g_2(0)]\,\}$
and $\{\,(x_1,\g_2(\tau_1))\,|\,x_1\in[\g_1(\tau_1),\g_1(\tau_2)]\,\},$ respectively. By a similar argument, we show that $w$ is also $\CC^{k,1}$ on all the other segments.

The estimates  in (\ref{xnn3.3}) follow from the ones in Lemmas \ref{l3.1} and \ref{p3.4}. \hfill$\Box$

{\bf Proof of Proposition \ref{p03.4}.}\,\,\,We divided $R(z,a,b)$ into a sum of small rectangles and apply Lemma \ref{p3.4} to paste the solutions together. \hfill$\Box$

To have  density results in Theorem \ref{t1.1},  we also need estimates of some (boundary) traces of the solutions.
For $\si\in(0,t_0),$ let
$$\b_\si(t)=\g(\si)-t\F\dot\g(\si)\qfq t\in(0,t_\si),$$ where $t_\si>0$ is such that $\b_\si(t_\si)\in\pl E(\g).$

\begin{pro}\label{p3.2} Let $f_0$ and $X$ be of class $\CC^{0,1}.$
Let $q_0$ be of class $\WW^{2,2}$ and  $q_1,$ $f$ be of class $\WW^{1,2},$ respectively. Then problem $(\ref{3.7})$ admits a unique solution  $w\in \WW^{2,2}$ with the data $(\ref{3.8}).$
Moreover, there is a $C>0,$ independent of solutions $w,$ such that
\be \|w\|^2_{\WW^{2,2}}+\|w_{x_2}\circ\b_\si\|^2_{\WW^{1,2}}\leq C(\|q_0\|^2_{\WW^{2,2}}+\|q_1\|^2_{\WW^{1,2}}+\|f\|_{\WW^{1,2}}),\label{xnnn3.3}\ee where $\WW^{i,2}=\WW^{i,2}(E(\g))$ for $1\leq i\leq2.$
\end{pro}

{\bf Proof}\,\,\, A similar argument as for Theorem \ref{p3.1} shows that  a unique solution $w\in\WW^{2,2}(E(\g))$ with the data  (\ref{3.8}) exists, and  the estimate
\be\|w\|^2_{\WW^{2,2}}\leq C(\|q_0\|^2_{\WW^{2,2}}+\|q_1\|^2_{\WW^{1,2}}+\|f\|_{\WW^{1,2}})\label{xxnn3.33}\ee holds.

Let $\b_\si(t)=(\b_{\si1}(t),\b_{\si2}(t)).$ Using  equation (\ref{3.7}), we have
\beq w_{x_2x_2}\circ\b_\si(t)&&=w_{x_2x_2}\circ\gamma\circ\g_2^{-1}\circ\b_{\si2}(t)+\int_{\g_1\circ\g_2^{-1}\circ\b_{\si2}(t)}^{\b_{\si1}(t)}[\eta(f,w)]_{x_2}(\zeta_1,\b_{\si2}(t))d\zeta_1,\nonumber\eeq which yields
\beq&&|w_{x_2x_2}\circ\b_\si(t)|^2\leq2|w_{x_2x_2}\circ\gamma\circ\g_2^{-1}\circ\b_{\si2}(t)|^2\nonumber\\
&&+2[\g_1(t_0)-\g_1(0)] C\int_{\g_1\circ\g_2^{-1}\circ\b_{\si2}(t)}^{\b_{\si1}(t)}(|f|^2+|\nabla f|^2+|w|^2+|\nabla w|^2+|\nabla^2w|^2)(\zeta_1, \b_{\si2}(t))d\zeta_1.\nonumber\eeq
Integrating the above inequality over $(0,t_\si)$ with respect to $t,$ we obtain
$$\|w_{x_2x_2}\circ\b_\si\|^2_{L^2}\leq C(\|f\|^2_{\WW^{1,2}}+\|q_0\|^2_{\WW^{2,2}}+\|q_1\|^2_{\WW^{1,2}}+\|w\|^2_{\WW^{2,2}}).$$
A similar computation shows that $\|\na w_{x_1}\circ\b_\si\|^2_{L^2},$  $\|\na w\circ\b_\si\|^2_{L^2},$ and $\|w\circ\b\|^2_{L^2}$ can be bounded also by the right hand side of the above inequality. Thus  estimate
(\ref{xnnn3.3}) follows from (\ref{xxnn3.33}). \hfill$\Box$

Let
\be\Ga(\g,w)=\sum_{j=0}^{1}\|\nabla^jw\circ\gamma\|^2_{L^2(0,t_0)}+\int_0^{t_0}[|w_{x_1x_1}\circ\gamma(t)|^2t+|w_{x_2x_2}
\circ\gamma(t)|^2(t_0-t)]dt.\label{g3.7}\ee
\begin{pro}\label{p3.3} Let $f_0$ and $X$ be of class $\CC^{0,1}.$ Then there are $0<c_1<c_2$ such that for all solutions $w\in\WW^{2,2}$ to problem $(\ref{3.7})$
\be c_1\Ga(\g,w)\leq\|f\|^2_{\WW^{1,2}}+\|w\|^2_{\WW^{2,2}}\leq c_2[\|f\|^2_{\WW^{1,2}}+\Ga(\g,w)],\label{xxnn3.5}\ee
$$\|w(\cdot,\g_2(0))\|^2_{\WW^{1,2}(\g_1(0),\g_1(t_0))}+\int_{\g_1(0)}^{\g_1(t_0)}|w_{x_1x_1}(x_1,\g_2(0))|^2(x_1-\g_1(0))dx_1\leq c_2[\|f\|^2_{\WW^{1,2}}+\Ga(\g,w)],$$
$$\|w(\g_1(t_0),\cdot)\|^2_{\WW^{1,2}(\g_2(t_0),\g_2(0))}+\int_{\g_2(t_0)}^{\g_2(0)}|w_{x_2x_2}(\g_1(t_0),x_2)|^2(x_2-\g_2(t_0))dx_2\leq c_2[\|f\|^2_{\WW^{1,2}}+\Ga(\g,w)],$$
 where $\WW^{i,2}=\WW^{i,2}(E(\g))$ for $1\leq i\leq2.$
\end{pro}

\begin{pro}\label{p03.3} Let $f_0$ and $X$ be of class $\CC^{0,1}.$ Then there is $C>0$ such that for all solutions $w\in\WW^{2,2}(R(z,a,b))$ to problem $(\ref{3.7})$
\be \|w\|^2_{\WW^{2,2}}\leq C(\|f\|^2_{\WW^{1,2}}+\|p_1\|^2_{\WW^{2,2}(z_1,z_1+a)}+\|p_2\|^2_{\WW^{2,2}(z_2,z_2+b)}).\label{0xxnn3.5}\ee
\end{pro}

The proofs of the above two propositions will complete from Lemmas \ref{l3.3} and \ref{p3.5} below  by an argument as  for Proposition \ref{p3.1}. We omit the details.

\begin{lem}\label{l3.3}  Let  $T>0$ be given.
There is $\varepsilon_T>0$ such that if $|\g(0)|\leq T$ and $\max\{\g_1(t_0)-\g_1(0),\g_2(0)-\g_2(t_0)\}<\varepsilon_T,$ Proposition $\ref{p3.3}$ holds.
\end{lem}

{\bf Proof}\,\,\, {\bf Step 1}\,\,\,Using (\ref{3.7}) we have
\beq w_{x_1x_1}(x)&&=w_{x_1x_1}\circ\gamma\circ\g_1^{-1}(x_1)+\int_{\g_2\circ\g_1^{-1}(x_1)}^{x_2}[\eta(f,w)]_{x_1}(x_1,\zeta_2)d\zeta_2,\nonumber\eeq which yields
$$|w_{x_1x_1}(x)|^2\leq2|w_{x_1x_1}\circ\gamma\circ\g_1^{-1}(x_1)|^2+2[x_2-\g_2\circ\g_1^{-1}(x_1)]\int_{\g_2\circ\g_1^{-1}(x_1)}^{\g_2(0)}|[\eta(f,w)]_{x_1}(x_1,\zeta_2)|^2d\zeta_2,$$ and
$$|w_{x_1x_1}\circ\gamma\circ\g_1^{-1}(x_1)|^2\leq2|w_{x_1x_1}(x)|^2+2[x_2-\g_2\circ\g_1^{-1}(x_1)]\int_{\g_2\circ\g_1^{-1}(x_1)}^{\g_2(0)}|[\eta(f,w)]_{x_1}(x_1,\zeta_2)|^2d\zeta_2,$$ respectively.
Integrating thee above two inequalities, first with respect to $x_2$ over $(\g_2\circ\g_1^{-1}(x_1),\g_2(0))$  and then with respect to $x_1$
 over $(\g_1(0),\g_1(t_0))$  respectively, we obtain
$$\|w_{x_1x_1}\|^2_{L^2}\leq2\si_{12}\int_0^{t_0}|w_{x_1x_1}\circ\gamma(t)|^2tdt+\varepsilon^2_TC_T(\|f\|^2_{\WW^{1,2}}+\|w\|^2_{\WW^{2,2}})$$ and
$$\si_{11}\int_0^{t_0}|w_{x_1x_1}\circ\gamma(t)|^2tdt\leq2\|w_{x_1x_1}\|^2_{L^2}+\varepsilon_T^2C_T(\|f\|^2_{\WW^{1,2}}+\|w\|^2_{\WW^{2,2}}),$$ where
$$\si_{11}=\inf_{t\in(0,t_0)}[\g_2(0)-\g_2(t)]\g_1'(t)/t,\quad\si_{12}=\sup_{t\in(0,t_0)}[\g_2(0)-\g_2(t)]\g_1'(t)/t,$$
$$C_T=\sup_{|x|\leq 2T}(1+f_0^2+|\nabla f_0|^2+|X|^2+|\nabla X|^2).$$
By similar arguments, we establish the following
$$ \|w_{x_2x_2}\|^2_{L^2}\leq2\si_{22}\int_0^{t_0}|w_{x_2x_2}\circ\gamma(t)|^2(t_0-t)dt+\varepsilon^2_TC_T(\|f\|^2_{\WW^{1,2}}+\|w\|^2_{\WW^{2,2}}),$$
$$\si_{21}\int_0^{t_0}|w_{x_2x_2}\circ\gamma(t)|^2(t_0-t)dt\leq2\|w_{x_2x_2}\|^2_{L^2}+\varepsilon^2_TC_T(\|f\|^2_{\WW^{1,2}}+\|w\|^2_{\WW^{2,2}}),$$ where
$$\si_{21}=\inf_{t\in(0,t_0)}[\g_1(t_0)-\g_1(t)][-\g_2'(t)]/(t_0-t),\quad\si_{22}=\sup_{t\in(0,t_0)}[\g_1(t_0)-\g_1(t)][-\g_2'(t)]/(t_0-t).$$

{\bf Step 2}\,\,\,As in Step 1, we have
\beq\|w_{x_1}\|^2_{L^2}&&\leq2\si_{12}\int_0^{t_0}|w_{x_1}\circ\gamma(t)|^2tdt+\varepsilon^2_TC_T(\|f\|^2_{L^2}+\|w\|^2_{\WW^{1,2}})\nonumber\\
&&\leq2\si_{12}t_0\|w_{x_1}\circ\gamma\|_{L^2(0,t_0)}^2+\varepsilon^2_TC_T(\|f\|^2_{L^2}+\|w\|^2_{\WW^{1,2}}),\nonumber\eeq
\be\si_{11}\int_0^{t_0}|w_{x_1}\circ\gamma(t)|^2tdt\leq2\|w_{x_1}\|^2_{L^2}+\varepsilon^2_TC_T(\|f\|^2_{L^2}+\|w\|^2_{\WW^{1,2}}).\label{g3.43}\ee
In addition, since
$$w_{x_1}\circ\gamma\circ\g_2^{-1}(x_2)=w_{x_1}(x)-\int_{\g_1\circ\g_2^{-1}(x_2)}^{x_1}w_{x_1x_1}(\zeta_1,x_2)d\zeta_1\qfq x_2\in(\g_2(t_0),\g_2(0)),$$
it follows that
\be\si_{21}\int_0^{t_0}|w_{x_1}\circ\gamma(t)|^2(t_0-t)dt\leq2\|w_{x_1}\|^2_{L^2}+\varepsilon_T^2\|w_{x_1x_1}\|^2_{L^2}. \label{g3.44}\ee
Combing (\ref{g3.43}) and (\ref{g3.44}), we have
\beq\min\{\si_{11},\si_{21}\}\|w_{x_1}\circ\gamma\|^2_{L^2(0,t_0)}&&\leq\frac{1}{t_0}(\si_{21}\int_0^{t_0/2}|w_{x_1}\circ\gamma(t)|^2(t_0-t)dt+\si_{11}\int_{t_0/2}^{t_0}|w_{x_1}\circ\gamma(t)|^2tdt)\nonumber\\
&&\leq\frac{1}{t_0}[4+\varepsilon_T^2(C_T+1)](\|f\|^2_{L^2}+\|w\|^2_{\WW^{2,2}}).\nonumber\eeq
By a similar computation, we obtain
$$\|w_{x_2}\|^2_{L^2}\leq2\si_{12}t_0\|w_{x_2}\circ\gamma\|^2_{L^2(0,t_0)}+C_T\varepsilon_T^2(\|f\|^2_{L^2}+\|w\|^2_{\WW^{1,2}}),$$
$$\min\{\si_{11},\si_{21}\}\|w_{x_2}\circ\gamma\|^2_{L^2(0,t_0)}\leq\frac{1}{t_0}[4+\varepsilon_T^2(C_T+1)](\|f\|^2_{L^2}+\|w\|^2_{\WW^{2,2}}),$$
$$ \|w\|^2_{L^2}\leq2\si_{12}t_0\|w\circ\gamma\|^2_{L^2(0,a)}+\varepsilon_T^2\|w_{x_2}\|^2_{L^2},$$
$$\min\{\si_{11},\si_{21}\}\|w\circ\gamma\|^2_{L^2(0,t_0)}\leq\frac{1}{t_0}[4+\varepsilon_T^2(C_T+1)]\|w\|^2_{\WW^{1,2}}.$$

{\bf Step 3}\,\,\,From Steps 1 and 2, we obtain
\beq&&[1-(4C_T+1)\varepsilon_T^2]\|w\|^2_{\WW^{2,2}}\leq2[(\si_{12}(1+t_0)+\si_{22}]\Ga(\g,w)+(4C_T+1)\varepsilon_T^2\|f\|^2_{W^{1,2}},\nonumber\eeq
when $\lam$ is small, and
$$\min\{\si_{11},\si_{21}\}\Ga(\g,w)\leq2\{2+C_T\varepsilon_T^2+\frac{3}{t_0}[4+(C_T+1)\varepsilon_T^2]\}(\|f\|^2_{\WW^{1,2}}+\|w\|^2_{\WW^{2,2}}),$$ respectively. Thus (\ref{xxnn3.5}) follows.

{\bf Step 4}\,\,\, We have
$$w_{x_1x_1}(x_1,\g_2(0))=w_{x_1x_1}\circ\g\circ\g_1^{-1}(x_1)+\int_{\g_2\circ\g_1^{-1}(x_1)}^{\g_2(0)}[\eta(f,w)]_{x_1}(x_1,\zeta_2)d\zeta_2,$$ which gives, by (\ref{xxnn3.5}),
\beq&&\int_{\g_1(0)}^{\g_1(t_0)}|w_{x_1x_1}(x_1,\g_2(0))|^2(x_1-\g_1(0))dx_1\leq2\int_0^{t_0}|w_{x_1x_1}\circ\g(t)|^2[\g_1(t)-\g_1(0)]dt\nonumber\\
&&\quad+C(\|f\|^2_{\WW^{1,2}}+\|w\|^2_{\WW^{2,2}})\leq C[\|f\|^2_{\WW^{1,2}}+\Ga(\g,w)].\nonumber\eeq A similar argument completes the proof of the third inequality in Proposition \ref{p3.3}.\hfill$\Box$

A similar argument yields the following.
\begin{lem}\label{p3.5} Let  $T>0$ be given.
There is $\varepsilon_T>0$ such that if $|z|\leq T$ and $0<\max\{a,b\}<\varepsilon_T,$ then Proposition $\ref{p03.3}$ holds.
\end{lem}

\subsection{ Regions $P_i(\b)$}
\hskip\parindent
Let $\b=(\b_1,\b_2):$ $[0,t_0]\rw\R^2$ be a noncharcteristic curve with $\b_1'(0)\b_2'(0)>0.$ We assume
\be\b_i'(t)>0\qfq t\in[0,t_0],\quad i=1,\,\,2.\label{xn3.4}\ee Otherwise, we consider the curve $z(t)=\b(-t+t_0).$
Set
\be P_1(\b)=\{\,(x_1,x_2)\,|\,\b_1\circ\b_2^{-1}(x_2)<x_1<\b_1(t_0),\,\,\b_2(0)<x_2<\b_2(t_0)\,\},\label{P_1}\ee
and consider the boundary data
\be w_{x_2}\circ\b(t)=p(t),\quad t\in(0,t_0);\quad w(x_1,\b_2(0))=p_1(x_1),\quad x_1\in(\b_1(0),\b_1(t_0)).\label{xnn3.11}\ee
Set
\be P_2(\b)=\{\,(x_1,x_2)\,|\,\b_1(0)<x_1<\b_1\circ\b_2^{-1}(x_2),\,\,\b_2(0)<x_2<\b_2(t_0)\,\},\label{P_2}\ee and consider the boundary data
\be w_{x_1}\circ\b(t)=p(t),\quad t\in(0,t_0);\quad w(\b_1(0),x_2)=p_2(x_2),\quad x_1\in(\b_1(0),\b_1(t_0)).\label{xnn3.12}\ee

By similar arguments for the region $E(\g),$ we establish Propositions \ref{p3.7}-\ref{p3.9} below. The details are omitted.

\begin{pro}\label{p3.7} Let the curve $\b$ be of class $\CC^{k-1,1}.$
  Let $p_1$ $($or $p_2$ $)$ be of class $\CC^{k,1}$ and let $p,$ $f$ be of class $\CC^{k-1,1}.$
Then problem $(\ref{3.7})$ admits a unique solution $w\in \CC^{k,1}(\overline{P_1(\b)})$ $($or $\CC^{k,1}(\overline{P_2(\b)})$ with the data $(\ref{xnn3.11})$ ( or (\ref{xnn3.12})) to satisfy
$$\|w\|_{\CC^{k,1}}\leq C(\|p\|_{\CC^{k-1,1}}+\|p_1\|_{\CC^{k,1}}+\|f\|_{\CC^{k-1,1}})(\mbox{or}\,\,(\|p\|_{\CC^{k-1,1}}+\|p_2\|_{\CC^{k,1}}+\|f\|_{\CC^{k-1,1}})).$$
\end{pro}

\begin{pro}\label{p3.8} Let the curve $\b$ be of class $\CC^{1}.$ Let $f_0$ and $X$ be of class $\CC^{0,1}.$ Let  $p_1$ $($or $p_2$ $)$ be of class $\WW^{2,2}$ and let $p,$ $f$ be of class $\WW^{1,2}.$
Then problem $(\ref{3.7})$ admits a unique solution $w\in \WW^{2,2}({P_1(\b)})$ $($or $\WW^{2,2}({P_2(\b)})$ with the data $(\ref{xnn3.11})$ $($or $(\ref{xnn3.12}))$ to satisfy
$$\|w\|_{\WW^{2,2}}\leq C(\|p\|_{\WW^{1,2}}+\|p_1\|_{\WW^{2,2}}+\|f\|_{\WW^{1,2}})(\mbox{or}\,\,(\|p\|_{\WW^{1,2}}+\|p_2\|_{\WW^{2,2}}+\|f\|_{\WW^{1,2}})).$$
\end{pro}

Let
$$ \Ga(P_i,w)=\int_0^{t_0}|p'(t)|^2(t_0-t)dt+\|p_i\|^2_{\WW^{1,2}}+\int_{\b_i(0)}^{\b_i(t_0)}|p_i''(x_i)|^2(x_i-z_i)]dx_i,\quad i=1,\,\,2.  $$

\begin{pro}\label{p3.9}Let the curve $\b$ be of class $\CC^{1}.$ Let $f_0$ and $X$ be of class $\CC^{0,1}.$ Then there are $0<c_1<c_2$ such that for all solutions $w\in \WW^{2,2}(P_i(\b))$ to problem $(\ref{3.7})$ with the corresponding boundary data satisfy
$$c_1\Ga(P_i,w)\leq\|w\|^2_{\WW^{2,2}}+\|f\|^2_{\WW^{1,2}}\leq c_2[\Ga(P_i,w)+\|f\|^2_{\WW^{1,2}}],$$
\beq &&c_1\|w|_{x_i=\b_i(t_0)}\|^2_{\WW^{2,2}}\leq \int_0^{t_0}|p'(t)|^2dt+\int_{\b_i(0)}^{\b_i(t_0)}|p_i''(x_1)|^2(x_i-\b_i(0))]dx_i+\|p_i\|^2_{\WW^{1,2}}+\|f\|^2_{\WW^{1,2}}\nonumber\\
&&\quad\leq c_2(\|w|_{x_i=\b_i(t_0)}\|^2_{\WW^{2,2}}+\int_{\b_i(0)}^{\b_i(t_0)}|p_i''(x_i)|^2(x_i-\b_i(0))]dx_i+\|p_i\|^2_{\WW^{1,2}}+\|f\|^2_{\WW^{1,2}}),\label{ga3.13}\eeq
for $i=1,$ and $2,$ respectively.
\end{pro}

\begin{rem} $(\ref{ga3.13})$ implies that $p\in\WW^{1,2}$ if and only if $w|_{x_i=\b_i(t_0)}\in\WW^{2,2}.$ However, the case of $p\notin\WW^{1,2}$ may happen.
\end{rem}

\subsection{Regions $\Xi_i(\b,\g)$}
\hskip\parindent Let $\g:$ $[0,t_1]\rw\R^2$ and $\b:$ $[0,t_0]\rw\R^2$ be two noncharacterstic curves  with $\g(0)=\b(0)$ such that
  $$\g_1(t_1)\leq\b_1(t_0),\quad\g_1'(t)>0,\quad \g_2'(t)<0,\quad\b_1'(t)>0,\quad\b_2'(t)>0$$ hold.  Set
\beq&&\overline{\Xi_1(\b,\g)}=\overline{P_1(\b)}\cup \overline{R(z,a,b)}\cup\overline{ E(\g)},\label{set3.15}\eeq
where $P_1(\b),$ $R(z,a,b),$ and $E(\g)$ are given in (\ref{P_1}), (\ref{R}), and (\ref{T(Z,a)}), respectively, with $z=(\b_1(t_0),\g_2(0)),$ $a=\g(t_1)-\b_1(t_0),$ and $b=\b_2(t_0)-\g_2(t_0).$
Consider the boundary data
\be w_{x_2}\circ\b(t)=p(t)\qfq t\in[0,t_0],\label{x3.25}\ee
\be w\circ\gamma(t)=q_0(t),\quad\<\nabla w,\F\dot\g\>\circ\gamma(t)=q_1(t)\qfq t\in(0,t_1),\label{x3.26}\ee where $\F$ is given by (\ref{f3.2}).

Let $\g:$ $[0,t_1]\rw\R^2$ and $\b:$ $[0,t_0]\rw\R^2$ be two noncharacterstic curves  with $\g(t_1)=\b(0)$ such that
$$\g_2(0)\geq\b_2(t_0),\quad\g_1'(t)>0,\quad\g_2'(t)<0,\quad \b_1'(t)>0,\quad\b_2'(t)>0$$ hold.
Set
\be\overline{\Xi_2(\b,\g)}=\overline{E(\g)}\cup\overline{R(z,a,b)}\cup\overline{P_2(\b)},\nonumber\ee where  $E(\g),$ $R(z,a,b),$ and $P_2(\b)$ are given in (\ref{T(Z,a)}),  (\ref{R}), and (\ref{P_2}), respectively,  with $z=(\g_1(t_1),\g_2(t_0)),$ $a=\b(t_0)-\g_1(t_1),$ and $b=\g_2(0)-\b_2(t_0).$
Consider  the data
\be w_{x_1}\circ\b(t)=p(t)\qfq t\in[0,t_0],\label{xx3.25}\ee
\be w\circ\gamma(t)=q_0(t),\quad\<\nabla w,\F\dot\g\>\circ\gamma(t)=q_1(t)\qfq t\in(0,t_1),\ee where $\F$ is given by (\ref{f3.2}).

We consider solvability of (\ref{3.7}) on $\Xi_1(\b,\g).$
To have a $\CC^{k,1}$ solution on $\overline{\Xi_1(\b,\g)},$ we need some kind of {\it compatibility conditions at the point $\g(0)=\b(0).$}
From Proposition \ref{p3.1}, problem (\ref{3.7}) admits a unique solution
$u\in \CC^{k,1}(\overline{E(\g)})$ with the data (\ref{x3.26}). From Proposition \ref{p3.7}, there is a unique solution $v\in \CC^{k,1}(\overline{P_1(\b}))$
to problem (\ref{3.7}) with the data
\be v_{x_2}\circ\b(t)=p(t),\quad t\in(0,t_0),\quad v(x_1,\b_2(0))=u(x_1,\b_2(0)),\quad x_1\in[\b_1(0),\b_1(t_0)].\label{3.27}\ee In terms of the uniqueness, if problem (\ref{3.7}) has a unique solution $w\in \CC^{k,1}(\overline{\Xi_1(\b,\g}))$
with the data (\ref{x3.25}) and (\ref{x3.26}) together, then
\be w(x)=\left\{\begin{array}{l}v(x)\qfq x\in \overline{P_1(\b)},\\
u(x)\qfq x\in\overline{E(\g)}.\end{array}\right.\label{3.28}\ee
Conversely, if we define $w$ by the formula (\ref{3.28}), then whether it is a $\CC^{k,1}$ solution to (\ref{3.7}) on $\Xi_1(\b,\g)$ depends on the $\CC^{k,1}$ regularity of $w$ at the point
$\b(0).$ Thus, {\it compatibility conditions} are something which can guarantee that $w$ is $\CC^{k,1}$ at $\g(0)=\b(0),$ that is
\be\nabla^ju\circ\g(0)=\nabla^jv\circ\b(0)\qfq 0\leq j\leq k.\label{xn3.10}\ee

The solution $u$ with the data (\ref{x3.26}) yields
\be\nabla u\circ\gamma(t)=\frac{1}{|\g'(t)|^2}(\g_1'(t)q_0'(t)+\g_2'(t)q_1(t),\,\g_2'(t)q_0'(t)-\g_1'(t)q_1(t))\label{3.29}\ee for $t\in[0,t_1].$
Using (\ref{3.7}) and (\ref{3.29}), we have
\beq u_{x_2x_1}\circ\gamma(t)&&=f\circ\gamma(t)+\frac{1}{|\g'(t)|^2}[\g_2'(t)X_1\circ\gamma(t)-\g_1'(t)X_2\circ\gamma(t)]q_1(t)\nonumber\\
&&\quad+f_0\circ\gamma(t)q_0(t)+\frac{1}{|\g'(t)|^2}[\g_1'(t)X_1\circ\gamma(t)+\g_2'(t)X_2\circ\gamma(t)]q_0'(t)\label{xxn3.14}\eeq  for $t\in(0,t_1).$   Next, differentiating the second component in
(\ref{3.29}) with respect to variable $t$ and using (\ref{xxn3.14}), we obtain
\beq &&u_{x_2x_2}\circ\gamma(t)=-\frac{\g_1'}{\g_2'}f\circ\gamma(t)-\frac{\g_1'}{\g_2'}f_0\circ\g q_0\nonumber\\
&&\quad-[\frac{2\<\g'',\g'\>}{|\g'|^4}+\frac{\g_1'}{|\g'|^2\g_2'}(\g_1'X_1\circ\g+\g_2'X_2\circ\g)-\frac{\g_2''}{|\g'|^2\g_2'}]q_0'+\frac{1}{|\g'|^2}q_0''\nonumber\\
&&\quad+[\frac{2\<\g'',\g'\>\g_1'}{|\g'|^4\g_2'}-\frac{\g_1'}{|\g'|^2\g_2'}(\g_2'X_1\circ\g-\g_1'X_2\circ\g)-\frac{\g_1''}{|\g'|^2\g_2'}]q_1-\frac{\g_1'}{|\g'|^2\g_2'}q_1'.\label{ga21}\eeq
By repeating the above procedure, we have shown that, for $1\leq j\leq k-1,$ there are $j$ order tensor fields  $A_{\a\b}(t),$ $A_\a^1(t),$ and $A_\a^0(t)$
such that
\beq \nabla^ju_{x_2}\circ\gamma(t)&&=\sum_{\a+\b\leq j-1}\pl_{x_1}^\a\pl_{x_2}^\b f\circ\gamma(t) A_{\a\b}(t)+\sum_{\a\leq j}q_1^{(\a)}(t)A^1_\a(t)\nonumber\\
&&\quad+\sum_{\a\leq j+1}q_0^{(j)}(t)A^0_\a(t)\qfq t\in[0,t_1].\label{x3.30}\eeq

Let $v\in \CC^{k,1}(\overline{P_1(\b}))$ be the solution to (\ref{3.7}) with the data (\ref{3.27}). Then
$$p'(t)=\<\nabla v_{x_2}(\b(t)),\,\dot\b(t)\>,\quad p''(t)=\<\nabla^2v_{x_2}(\b(t)),\,\dot\b(t)\otimes\dot\b(t)\>+\<\nabla v_{x_2}(\b(t)),\,\ddot\b(t)\>$$ for $t\in[0,t_0].$
Some computations show that
\beq p^{(l)}(t)&&=\<\nabla^lv_{x_2}(\b(t)),\,\dot\b(t)\otimes\cdots\otimes\dot\b(t)\>\nonumber\\
&&+\sum_{j_1+\cdots+j_i=l,\,1\leq i\leq l-1}a_{j_1\cdots j_i}\<\nabla^iv_{x_2}(\b(t)),\,\,\b^{(j_1)}(t)\otimes\cdots\otimes\b^{(j_i)}(t)\>
\label{x3.31}\eeq for $t\in[0,t_0],$ and $1\leq l\leq k,$ where $a_{j_1\cdots j_i}$ are positive integers. Then assumption (\ref{xn3.10}) is stated as the following.

\begin{dfn}\label{d3.1} Let the curves $\b$ and $\g$ be of class $\CC^{k,1}.$ Let  $q_0$ be of class $\CC^{k,1}$ and $p,$  $q_1,$  $f$  of class $\CC^{k-1,1},$ respectively.
It is said that  the {\it $k$th order compatibility conditions} hold  at $\g(0)=\b(0)$ if $|\g'(0)|^2p(0)=\g_2'(0)q_0'(0)-\g_1'(0)q_1(0)$ and
\beq  p^{(l)}(0)&&=\<\nabla^lu_{x_2}\circ\g(0),\,\dot\b(0)\otimes\cdots\otimes\dot\b(0)\>\nonumber\\
&&+\sum_{j_1+\cdots+j_i=l,\,1\leq i\leq l-1}a_{j_1\cdots j_i}\<\nabla^iu_{x_2}\circ\g(0),\,\,\b^{(j_1)}(0)\otimes\cdots\otimes\b^{(j_i)}(0)\>
\label{3.32}\eeq for  $1\leq l\leq k-1,$ where $\nabla^i u_{x_2}\circ\gamma(0)$ and $a_{j_1\cdots j_i}$  are given in $(\ref{x3.30})$ and $(\ref{x3.31}),$ respectively.
\end{dfn}

\begin{pro}\label{p3.10}Let the curves $\b$ and $\g$ be of class $\CC^{k,1}.$ Let  $q_0$ be of class $\CC^{k,1}$ and $p,$  $q_1,$  $f$  of class $\CC^{k-1,1},$ respectively. If $k\geq1,$ we assume that the $k$th order compatibility conditions hold  at $\g(0)=\b(0).$ Then problem $(\ref{3.7})$ admits a unique solution $w\in \CC^{k,1}(\overline{\Xi_1(\b,\g)})$ with the data $(\ref{x3.25})$ and $(\ref{x3.26}).$ Moreover, the following estimates hold
$$\|w\|_{\CC^{k,1}}\leq C(\|p\|_{\CC^{k-1,1}}+\|q_0\|_{\CC^{k,1}}+\|q_1\|_{\CC^{k-1,1}}+\|f\|_{\CC^{k-1,1}}).$$
\end{pro}

{\bf Proof}\,\,\,The uniqueness and the estimate follows from Propositions \ref{p3.1}, \ref{p03.4}, and \ref{p3.7}. It is remaining to show the existence. Let $u$ and $v$ be given in (\ref{3.28}) with the corresponding boundary date. Let $h$ be the solution to $(\ref{3.7})$ on $R(z,a,b)$ with the data
$$h(x_1,\g_2(0))=u(x_1,\g_2(0))\qfq x_1\in[\b_1(t_0),\g_1(t_1)],$$
$$h(\b_1(t_0),x_2)=v(\b_1(t_0),x_2)\qfq x_2\in[\g_2(0),\b_2(t_0)],$$ where $R(z,a,b)$ is given in (\ref{set3.15}).
We now define
$$w(x)=\left\{\begin{array}{l}v\qfq x\in\overline{P_1(\b)},\\
u\qfq x\in\overline{E(\g)},\\
h\qfq x\in\overline{R(z,a,b)}.\end{array}\right.$$ Then $w$ is a solution to (\ref{3.7}) with the data (\ref{x3.25}) and (\ref{x3.26}). Next we shall show $w\in\CC^{k,1}(\overline{\Xi_1(\b,\g)}).$

We proceed by induction in $k\geq0.$ The definition of $w$ guarantees $w\in\CC^{0,1}(\overline{\Xi_1(\b,\g)}).$
Let $w\in\CC^{k,1}(\overline{\Xi_1(\b,\g)}).$ Next we show that  the $k+1$th order compatibility conditions imply $w\in\CC^{k+1,1}(\overline{\Xi_1(\b,\g)}).$
For this purpose it is enough to show that $w$ is $\\C^{k+1}$ on the segments
$$\vartheta=\{\,(x_1,\b_2(0)),\,\,(x_2,\b_1(t_0))\,|\,x_1\in[\b_1(0),\g_1(t_0)],\,\,x_2\in[\g_2(0),\b_2(t_0)]\,\}.$$

By the induction assumptions, we have
\be \pl_{x_1}^i\pl_{x_2}^j v(x_1,\b_2(0))=\pl_{x_1}^i\pl_{x_2}^j u(x_1,\b_2(0))\qfq \b_1(0)\leq x_1\leq\b_1(t_0),\label{x3.28}\ee for $0\leq i+j\leq k.$
Next we show that (\ref{x3.28}) are true with
$i+j=k+1.$ Since $v(x_1,\b_2(0))=u(x_1,\b_2(0))$ for all $x_1\in[\b_1(0),\b_1(t_0)],$ it follows that
$$\pl_{x_1}^{k+1} v(x_1,\b_2(0))=\pl_{x_1}^{k+1}u(x_1, \b_2(0))\quad\mbox{for all}\quad x_1\in[\b_1(0),\b_1(t_0)].$$
Let $i+j=k+1$ with $j\geq1.$ If $i\geq1,$ then $j=k+1-i\leq k$ and, by the induction assumptions,
$$\pl^j_{x_2}v(x_1,\b_2(0))=\pl^j_{x_2}u(x_1,\b_2(0))\quad\mbox{for all}\quad x_1\in[\b_1(0),\b_1(t_0)],$$ which yield
\be\pl_{x_1}^i\pl^j_{x_2}v(x_1,\b_2(0))=\pl_{x_1}^i\pl^j_{x_2}u(x_1,\b_2(0))\quad\mbox{for all}\quad x_1\in[\b_1(0),\b_1(t_0)].\label{3.39}\ee Next we check the case of $i=0$ and
$j=k+1.$

Using (\ref{3.7}), we have
\beq\Big(\pl^{k+1}_{x_2}v(x_1,\b_2(0))\Big)_{x_1}&&=\pl_{x_2}^k(v_{x_1x_2})(x_1,\b_2(0))=\pl_{x_2}^k[f+f_0v+X_1v_{x_1}+X_2v_{x_2}](x_1,\b_2(0))\nonumber\\
&&=X_2(x_1,\b_2(0))\pl_{x_2}^{k+1}v(x_1,\b_2(0))+\pl_{x_2}^k[f+f_0v+X_1v_{x_1}](x_1,\b_2(0))\nonumber\\
&&\quad+[\sum_{i=1}^{k}C_k^i\pl_{x_2}^iX_2\pl_{x_2}^{k-i+1}v](x_1,\b_2(0)).\label{x3.40}\eeq
Let
$$\rho(x_1)=\pl_{x_2}^k[f+f_0v+X_1v_{x_1}](x_1,\b_2(0))+[\sum_{i=1}^{k}C_k^i\pl_{x_2}^iX_2\pl_{x_2}^{k-i+1}v](x_1,\b_2(0))$$ for $x_1\in[\b_1(0),\b_1(t_0)].$
It follows from (\ref{x3.40}) that $\tau(x_1)=\pl^{k+1}_{x_2}v(x_1,\b_2(0))$ is the solution to problem
\be\left\{\begin{array}{l}\tau'(x_1)=X_2(x_1,\b_2(0))\tau(x_1)+\rho(x_1)\qfq x_1\in[\b_1(0),\b_1(t_0)],\\
\tau(\b_1(0))=\pl^{k+1}_{x_2}v(\b(0)).\end{array}\right.\label{3.41}\ee
Moreover, the induction assumptions, $w\in \CC^{k,1}(\overline{\Xi_1(\b,\g))}),$ yield
$$\<\nabla^iv_{x_2}(z),\,\b^{(j_1)}(0)\otimes\cdots\b^{(j_i)}(0)\>=\<\nabla^iu_{x_2}(z),\,\b^{(j_1)}(0)\otimes\cdots\b^{(j_i)}(0)\>$$ for $j_1+\cdots+j_i=l,$ $1\leq i\leq l-1,$ and $1\leq l\leq k.$
Then the $k+1$th order compatibility conditions imply
$$ \<\nabla^{k}v_{x_2}\circ\b(0),\,\dot\b(0)\otimes\cdots\otimes\dot\b(0)\>=\<\nabla^{k}u_{x_2}\circ\g(0),\,\dot\b(0)\otimes\cdots\otimes\dot\b(0)\>.$$ Using (\ref{3.39}) and $\b_2'(0)>0,$ we obtain
\be\pl_{x_2}^{k+1}u\circ\g(0)=\pl_{x_2}^{k+1}v\circ\b(0).\label{inital}\ee In addition, it follows from the induction assumptions and (\ref{3.39}) that
$$\rho(x_1)=\pl_{x_2}^k[f+f_0u+X_1u_{x_1}](x_1,\b_2(0))+[\sum_{i=1}^{k}C_k^i\pl_{x_2}^iX_2\pl_{x_2}^{k-i+1}u](x_1,\b_2(0)),$$ for $ x_1\in[\b_1(0),\b_1(t_0)].$ By a similar computation as in (\ref{x3.40}), $\pl^{k+1}_{x_2}u(x_1,\b_2(0))$ is
also a solution to problem (\ref{3.41}) with the same initial date (\ref{inital}). The uniqueness of solutions of problem (\ref{3.41}) yields
$$\pl^{k+1}_{x_2}v(x_1,\b_2(0)) =\pl^{k+1}_{x_2}u(x_1,\b_2(0)),\quad x_1\in[\b_1(0),\b_1(t_1)].$$
Thus $w$ is $\CC^{k+1}$ on the segment
$$\{\,(x_1,\b_2(0))\,|\,x_1\in[\b_1(0),\b_1(t_0)]\,\}.$$
A similar argument shows that $w$ is $\CC^{k+1}$ on the rest of $\vartheta.$ The induction is complete.
\hfill$\Box$

By similar arguments, we have  Propositions \ref{p3.11}-\ref{np3.13} below. The details are omitted.

\begin{pro}\label{p3.11} Let the curves $\b$ and $\g$ be of class $\CC^{1}.$ Let $f_0$ and $X$ be of class $\CC^{0,1}.$ Let   $q_0$ be of class $\WW^{2,2}$ and $p,$ $q_1,$  $f$  of class $\WW^{1,2},$ respectively,
such that the $1$th order compatibility conditions hold true at $\g(0).$ Then problem $(\ref{3.7})$ admits a unique solution $w\in \WW^{2,2}(\Xi_1(\b,\g))$ with the data $(\ref{x3.25})$ and $(\ref{x3.26}).$ Moreover, the following estimates hold
$$\|w\|_{\WW^{2,2}}\leq C(\|p\|_{\WW^{1,2}}+\|q_0\|_{\WW^{2,2}}+\|q_1\|_{\WW^{1,2}}+\|f\|_{\WW^{1,2}}).$$
\end{pro}

Let
\be\Ga_i(\b,w)=\int_0^{t_0}|[w_{x_i}\circ\b(s)]'|^2(t_0-s)ds\qfq s\in(0,t_0),\quad i=1,\,2.\label{gam3.27}\ee

\begin{pro}\label{np3.12}Let the curves $\b$ and $\g$ be of class $\CC^{1}.$ Let $f_0$ and $X$ be of class $\CC^{0,1}.$ Then there are $0<c_1<c_2$ such that for all solutions $w\in \WW^{2,2}(\Xi_1(\b,\g))$ to problem $(\ref{3.7})$
$$ c_1[\Ga(\g,w)+\Ga_2(\b,w)]\leq\|w\|^2_{\WW^{2,2}}+\|f\|^2_{\WW^{1,2}}\leq c_2[\Ga(\g,w)+\Ga_2(\b,w)+\|f\|^2_{\WW^{1,2}}],$$ where $\Ga(\g,w)$ is given in
$(\ref{g3.7}).$
\end{pro}

\begin{pro}\label{np3.13} The corresponding results as in Propositions $\ref{p3.10},$ $\ref{p3.11},$ and $\ref{np3.12}$ hold where $\Xi_1(\b,\g)$ and $\Ga_2(\b,w)$ are
 replaced with $\Xi_2(\b,\g)$ and $\Ga_1(\b,w),$ respectively.
\end{pro}

\subsection{Region $\Phi(\b,\g,\hat\b)$}
\hskip\parindent Let $\b:\,\,[0,t_0]\rw\R^2,$ $\g:\,\,[0,t_1]\rw\R^2,$ and $\hat\b:\,\,[0,t_2]\rw\R^2$  be noncharacteristic curves  with $\b(0)=\g(0)$ and $\g(t_1)=\hat\b(0)$ such that
$$\g_1(t_1)\geq\b_1(t_0),\quad\g_1'(t)>0,\quad\g_2'(t)<0,\quad\b_1'(t)>0,\quad\b_2'(t)>0,$$
$$\g_2(t_1)\geq\hat\b_2(t_2),\quad\hat\b_1'(t)>0,\quad\hat\b_2'(t)>0.$$
We define
\beq\overline{\Phi(\b,\g,\hat{\beta})}=\overline{\Xi_1(\b,\g)}\cup\overline{R(z,a,b)}\cup\overline{P_2(\hat\b)},\nonumber\eeq where $\Xi_1(\b,\g),$ $R(z,a,b),$ $P_2(\hat\b)$ are given in (\ref{set3.15}), (\ref{R}), and (\ref{P_2}), respectively, with $z=(\g_1(t_1),\hat\b_2(t_2)),$ $a=\hat\b_1(t_2)-\g_1(t_1),$ and $b=\b_2(t_0)-\hat\b(t_2).$
Consider the boundary data
\be w_{x_2}\circ\b(t)=p_1(t),\quad t\in[0,t_0],\quad w_{x_1}\circ\hat\b(t)=p_2(t),\quad t\in(0,t_2),\label{xx3.17}\ee
\be w\circ\gamma(t)=q_0(t),\quad\<\na w,\F\dot\g\>\circ\gamma(t)=q_1(t)\qfq t\in(0,t_1).\label{xx3.18}\ee

By similar arguments as for $\Xi_1(\b,\g),$ we have Propositions \ref{t3.1}-\ref{t3.3} below. The details are omitted.

\begin{pro}\label{t3.1} Let the curves $\b,$ $\g,$ and $\hat\b$ be of class $\CC^{k,1}.$ Let $q_0$ be of class $\CC^{k,1},$ and  $p_1,$ $p_2,$  $q_1,$ $f$ of class $\CC^{k-1,1}$
such that the $k$th order compatibility conditions hold true at $\g(0)$ and $\g(t_1),$ respectively. Then problem $(\ref{3.7})$ admits a unique solution
$w\in \CC^{k,1}(\overline{\Phi(\b,\g,\hat\b)})$ with the data $(\ref{xx3.17})$ and $(\ref{xx3.18}).$
Moreover, the following estimates hold
$$\|w\|^2_{\CC^{k,1}}\leq C(\|p_1\|^2_{\CC^{k-1,1}}+\|p_2\|^2_{\CC^{k-1,1}}+\|q_0\|^2_{\CC^{k,1}}+\|q_1\|^2_{\CC^{k-1,1}}+\|f\|^2_{\CC^{k-1,1}}).$$
\end{pro}

\begin{pro}\label{t3.2}Let the curves $\b,$ $\g,$ and $\hat\b$ be of class $\CC^{1}.$ Let $f_0$ and $X$ be of class $\CC^{0,1}.$ Let  $q_0$ be of class $\WW^{2,2},$ and $p_1,$ $p_2,$   $q_1,$ $f$ of class $\WW^{1,2},$
such that the $1$th order compatibility conditions hold true at $\g(0)$ and $\g(t_1),$ respectively. Then problem $(\ref{3.7})$ admits a unique solution
$w\in \WW^{2,2}({\Phi(\b,\g,\hat\b)})$ with the data $(\ref{xx3.17})$ and $(\ref{xx3.18}).$
Moreover, the following estimates hold
$$\|w\|^2_{\WW^{2,2}}\leq C(\|p_1\|^2_{\WW^{1,2}}+\|p_2\|^2_{\WW^{1,2}}+\|q_0\|^2_{\WW^{2,2}}+\|q_1\|^2_{\WW^{1,2}}+\|f\|^2_{\WW^{1,2}}).$$
\end{pro}

\begin{pro}\label{t3.3}
Let the curves $\b,$ $\g,$ and $\hat\b$ be of class $\CC^{1}.$ Let $f_0$ and $X$ be of class $\CC^{0,1}.$ Then there are $0<c_1<c_2$ such that for all solutions $w\in \WW^{2,2}(\Phi(\b,\g,\hat\b))$ to problem $(\ref{3.7})$
$$ c_1[\Ga(\g,w)+\Ga_1(\hat\b,w)+\Ga_2(\b,w)]\leq\|w\|^2_{\WW^{2,2}}+\|f\|^2_{\WW^{1,2}}\leq c_2[\Ga(\g,w)+\Ga_1(\hat\b,w)+\Ga_2(\b,w)+\|f\|^2_{\WW^{1,2}}],$$ where $\Ga(\g,w),$ $\Ga_1(\hat\b,w),$ and $\Ga_2(\b,w)$ are given in
$(\ref{g3.7})$ and $(\ref{gam3.27}),$ respectively.
\end{pro}

\setcounter{equation}{0}
\section{Solvability for Hyperbolic Surfaces}\label{s4}
\def\theequation{4.\arabic{equation}}
\hskip\parindent   Let $M\subset\R^3$ be a hyperbolic surface with  the normal field $\n$ and let $\Om\subset M$ be a noncharacteristic region, where
$$  \Om=\{\,\a(t,s)\,|\,(t,s)\in(0,a)\times(0,b)\,\}.$$
We consider solvability of problem under appropriate part boundary data
\be \<D^2w,Q^*\Pi\>=f+f_0w+X(w)\qfq x\in \Om,\label{3.1}\ee where $f_0$ is a function on $M$ and $X\in T(M)$ is a vector field on $M.$ Clearly, equation (\ref{2.17}) takes the form of (\ref{3.1}).

To set up boundary data, we consider some boundary operators. Let $x\in\pl \Om$ be given. $\mu\in T_xM$ with $|\mu|=1$ is said to be the {\it noncharacteristic normal} outside $\Om$ if there is a curve
$\zeta:$ $(0,\varepsilon)\rw \Om$ such that
$$\zeta(0)=x,\quad \zeta'(0)=-\mu,\quad\Pi(\mu,X)=0\qfq X\in T_x(\pl \Om).$$ Let $\mu$ be the the noncharacteristic normal field along $\pl \Om.$
Let the linear operator $Q:$ $T_xM\rw T_xM$ be given in (\ref{1.3n}) for $x\in M.$ Recall that the shape operator $\nabla\n:$ $T_xM\rw T_xM$ is defined by $\nabla\n X=\nabla_X\n(x)$ for $X\in T_xM.$
 We define boundary operators $\T_i:$ $T_xM\rw T_xM$ by
\be \T_iX=\frac{1}{2}\Big[X+(-1)^i\chi(\mu,X)\rho(X)Q\nabla\n X\qfq X\in T_xM,\quad i=1,\,\,2,\label{xn4.14}\ee where
\be\chi(\mu,X)=\sign\det\Big(\mu,X,\n\Big),\quad \varrho(X)=\frac{1}{\sqrt{-\kappa}}\sign\Pi(X,X),\label{rho4.3}\ee and  $\sign$ is the sign function.

We shall consider the part boundary data
\be\<Dw,\T_2\a_s\>\circ\a(0,s)=p_1(s),\quad \<Dw,\T_2\a_s\>\circ\a(a,s)=p_2(s)\qfq s\in(0,b),\label{4.3}\ee
\be w\circ\a(t,0)=q_0(t),\quad\frac{1}{\sqrt{2}}\<Dw,(\T_2-\T_1)\a_t\>\circ\a(t,0)=q_1(t)\qfq t\in(0,a).\label{x1}\ee

To have a smooth solution, we need some kind of {\it compatibility conditions} as follows.

Let $A$ and $B$ be $k$th order and $m$th order tensor fields on $M,$ respectively, with $k\geq m.$ We define $A(\ii-)B$ to be a $(k-m)$th order tensor field by
\be A(\ii-)B(X_1,\cdots,X_{k-m})=\<\ii_{X_{k-m}}\cdots \ii_{X_{1}}A,B\>(x)\qfq x\in M,\label{4.6}\ee where $X_1,$ $\cdots,$ $X_{k-m}$ are vector fields on $M.$

For convenience, we assume that
$$|\a_t(t,0)|=1\qfq t\in[0,a].$$ Then $Q\a_t,$ $\a_t$ forms an orthonormbal basis of $T_{\a(t,0)}M$  with the positive orientation for all $t\in[0,a]$ and
$$Q\na\n\a_t=\Pi(\a_t,\a_t)Q\a_t-\Pi(\a_t,Q\a_t)\a_t\qfq t\in[0,a].$$
Let $k\geq1$ be an integer. First, we assume that
$w$ is a $\CC^{k,1}$ solution to (\ref{3.1}) in a neighborhood of the curve $\a(t,0)$ with the data (\ref{x1}). Then
$$Dw(\a(t,0))=B_1(t)q_0'(t)+C_0(t)q_1(t)\qfq t\in[0,a],$$ where
$$B_1(t)=[\a_t+\frac{\Pi(\a_t,Q\a_t)}{\Pi(\a_t,\a_t)}Q\a_t],\quad C_0(t)=\frac{\sqrt{2}}{\varrho(\a_t)\Pi(\a_t,\a_t)}Q\a_t,$$ are vector fields along the curve $\a(t,0),$ from which we obtain
\beq D_{\a_t} Dw(\a(t,0))&&=D_{\a_t}B_1q_0'(t)+B_1(t)q_0''(t)+D_{\a_t}C_0q_1(t)+C_0(t)q_1'(t).\nonumber\eeq  Using
(\ref{3.1}) and  the above formula, we compute along the curve $\a(t,0)$ to have
\beq &&D^2w(Q\a_t,Q\a_t)\Pi(\a_t,\a_t)=f+f_0w+\<Dw,X\>-\<D_{\a_t} Dw,\a_t\>\Pi(Q\a_t,Q\a_t)\nonumber\\
&&\quad+2\<D_{\a_t}Dw,Q\a_t\>\Pi(Q\a_t,\a_t)\nonumber\\
&&=f+f_0q_0(t)+[\<X,B_1(t)\>+\<D_{\a_t}B_1,Z(t)\>]q_0'(t)+\<B_1(t),Z(t)\>q_0''(t)\nonumber\\
&&+[\<X,C_0(t)\>+\<D_{\a_t}C_0,Z(t)\>]q_1(t)+\<C_0(t), Z(t)q_1'(t)\qfq t\in[0,a],\label{xnn4.7}\eeq where
$$Z(t)=2\Pi(Q\a_t,\a_t)Q\a_t-\Pi(Q\a_t,Q\a_t)\a_t.$$
Since $\Pi(\a_t,\a_t)\not=0$ for all $t\in[0,a],$ we have obtained two order tensor fields, $A^2(t),$ $B^2_i(t),$ and $C^2_i(t),$ that are given by $f_0,$ $X,$ $\Pi,$
$Q\a_t,$ $\a_t,$ and their differentials, such that
$$D^2w(\a(t,0))=A^2(t)f+\sum_{i=0}^2B^2_i(t)q_0^{(i)}(t)+\sum_{i=0}^1C^2_i(t)q_1^{(i)}(t)\qfq t\in[0,a].$$
By repeating the above procedure, we obtain $(k+i)$th order tensors fields $A_i^{k+i}(t),$ and $k$th order tensor fields $B_i^k(t),$ $C_i^k(t),$ such that
$$D^kw(\a(t,0))=\Q_k(q_0,q_1,f)(t)\qfq t\in[0,a],$$ where
\be \Q_k(q_0,q_1,f)(t)=\sum_{i=0}^{k-2}A_i^{k+i}(t)(\ii-)D^if(\a(t,0))+\sum_{i=0}^kB_i^k(t)q_0^{(i)}(t)+\sum_{i=0}^{k-1}C^k_i(t)q_1^{(i)}(t)\label{xn4.7}\ee for $t\in[0,a]$ and $k\geq2,$ where
``$(\ii-)$" is defined in (\ref{4.6}).

\begin{dfn}\label{d4.2} Let  $q_0$ be of class $\CC^{k,1},$ and $p_1,$ $ p_2,$ $q_1,$  $f$ of class $\CC^{k-1,1}$  to be said to satisfy
the $k$th order compatibility conditions at $\a(0,0)$ and $\a(a,0)$ if
\be p_j(t_j)=\<B_1(t_j),\T_2\a_s\>q_0'(t_j)+\<C_0(t_j),\T_2\a_s\>q_1(t_j),\label{cn}\ee
\beq p_j^{(l)}(0)&&=\<\Q_l(q_0,q_1,f)(t_j),\,\dot\gamma_j(0)\otimes\cdots\otimes\dot\gamma_j(0)\>\nonumber\\
&&+\sum_{j_1+\cdots+j_i=l,\,1\leq i\leq l-1}a_{j_1\cdots j_i}\<\Q_i(q_0,q_1,f)(t_j),\,\,\gamma^{(j_1)}_j(0)\otimes\cdots\otimes\gamma^{(j_i)}_j(0)\>
\label{xn4.9}\eeq for $1\leq l\leq k-1,$ where $a_{j_1\cdots j_i}$ are positive integers given in $(\ref{x3.31}),$ $j=1,$ $2,$ $\gamma_1(s)=\a(0,s),$ $\gamma_2(s)=
\a(a,s),$ $t_1=0,$ and $t_2=a.$
\end{dfn}

Our main task in this section is to establish the following.

\begin{thm}\label{t4.1} Let $\Om$ be a noncharacteristic region of class $\CC^{m+2,1}$ and let $f_0$ and $X$ be of class $\CC^{m-1,1}.$
Let  $q_0$ be of class $\CC^{m,1},$ and $p_1, $ $p_2,$ $q_1,$ $f$ be of $\CC^{m-1,1},$ respectively. If $m\geq1,$ we assume that the $m$th compatibility conditions holds.
Then there is a unique solution $w\in \CC^{m,1}(\overline\Om)$ to problem $(\ref{3.1})$ with the data $(\ref{4.3})$ and $(\ref{x1})$ satisfying
\beq\|w\|_{\CC^{m,1}(\overline{\Om})}&&\leq C(\|q_1\|_{\CC^{m-1,1}[0,a]}+\|q_0\|_{\CC^{m,1}[0,a]}+\|p_1\|_{\CC^{m-1,1}[0,b]}\nonumber\\
&&\quad+\|p_2\|_{\CC^{m-1,1}[0,b])}+\|f\|_{\CC^{m-1,1}(\overline{\Om})}).\label{xn4.10}\eeq
\end{thm}

\begin{rem}If $p_1,$ $ p_2\in \CC_0^{m-1,1}(0,b),$ $q_0\in\CC_0^{m,1}(0,a),$ $q_1\in \CC_0^{m-1,1}(0,a),$ and $f\in \CC_0^{m-1,1}(\Om)$  for an integer $m\geq0,$ then   the $m$th order compatibility conditions are clearly true.
\end{rem}

\begin{thm}\label{t4.2} Let $\Om$ be a noncharacteristic region of class $\CC^{2,1}$ and let $f_0$ and $X$ be of class $\CC^{0,1}.$
 Let $q_0$ be of class $\WW^{2,2},$ and $p_1,$ $p_2,$ $q_1,$   $f$  of class $\WW^{1,2}$  to satisfy the $1$th order compatibility conditions.
Then there is a unique solution $w\in \WW^{2,2}(\Om)$ to problem $(\ref{3.1})$ with the data $(\ref{4.3})$ and $(\ref{x1}).$
Moreover, there is  $C>0,$ in dependent of solution $w,$ such that
\beq \|w\|^2_{\WW^{2,2}(\Om)}&&\leq C(\|q_0\|^2_{\WW^{2,2}(0,a)}+\|q_1\|^2_{\WW^{1,2}(0,a)}+\|p_1\|^2_{\WW^{1,2}(0,b)}\nonumber\\
&&\quad+\|p_2\|^2_{\WW^{1,2}(0,b)}+\|f\|_{\WW^{1,2}(\Om)}).\label{xn4.12}\eeq
\end{thm}

 We define
\beq\Ga(\Om,w)&&=\int_0^b(|p_1'(s)|^2+|p_2'(s)|^2)(b-s)ds+\Ga(\a(\cdot,0),w),\label{GaOm}\eeq where $p_1,$ $p_2$ are given in (\ref{4.3}), and
\beq\Ga(\a(\cdot,0),w)&&=\sum_{j=0}^1\|\na^jw\circ\a(\cdot,0)\|^2_{L^2(0,a)}\nonumber\\
&&\quad+\int_0^{a}[|D^2w(\T_1\a_t,\T_1\a_t)|^2t+|D^2w(\T_2\a_t,\T_2\a_t)|^2(a-t)]dt.\quad\label{ga4.13}\eeq

\begin{thm}\label{t4.3}
Let $\Om$ be a noncharacteristic region of class $\CC^{2,1}$ and let $f_0$ and $X$ be of class $\CC^{0,1}.$ Then there are $0<c_1<c_2$ such that for all solutions $w\in \WW^{2,2}(\Om)$ to problem $(\ref{3.1})$
\be c_1\Ga(\Om,w)\leq\|w\|^2_{\WW^{2,2}(\Om)}+\|f\|^2_{\WW^{1,2}(\Om)}\leq c_2(\|f\|^2_{\WW^{1,2}(\Om)}+\Ga(\Om,w)).\label{g4.19}\ee
\end{thm}

Next, we assume that $f=0$ to consider problem
\be \<D^2w,Q^*\Pi\>=f_0w+X(w)\qfq x\in \Om.\label{3xxn.1}\ee
Denote by $\Upsilon(\Om)$ all the solutions $w\in\WW^{2,2}(\Om)$  to problem (\ref{3xxn.1}).
For $w\in\Upsilon(\Om),$ we let
\beq \Ga(w)&&=\int_0^b(|p_1'(s)|^2+|p_2'(s)|^2)ds+\|q_0\|^2_{\WW^{2,2}(0,a)}+\|q_1\|^2_{\WW^{1,2}(0,a)},\nonumber\eeq where $p_1,$ $p_2,$ $q_0,$ and $q_1$ are given in (\ref{4.3}) and (\ref{x1}), respectively. We define
$$\H(\Om)=\{\,w\in\Upsilon(\Om)\,\mbox{with the $1$th order compatibility conditions}\,|\,\,\Ga(w)<\infty\,\}.$$

\begin{thm}\label{t4.4}Let $\Om$ be a noncharacteristic region of class $\CC^{2,1}$ and $X$ of class $\CC^{0,1}.$  For each $w\in\Upsilon(\Om),$  there exists a sequence $w_n\in\H(\Om)$ such that
$$\lim_{n\rw\infty}\|w_n-w\|_{\WW^{2,2}(\Om)}=0.$$
\end{thm}

The remains of this section is devoted to the proofs of Theorems \ref{t4.1}-\ref{t4.4}. The proofs of Theorems \ref{t4.1}-\ref{t4.2}  and \ref{t4.3}-\ref{t4.4} are given after Lemma \ref{l4.5} and
Lemma \ref{l4.7}, respectively.\\

We shall solve (\ref{3.1}) locally in asymptotic coordinate systems and then paste the local solutions together.  A chart $\psi(p)=(x_1,x_2)$ on $M$ is said to be an {\it asymptotic coordinate system} if
\be\Pi(\pl x_1,\pl x_1)=\Pi(\pl x_2,\pl x_2)=0.\label{3.4}\ee
Let $p\in M.$  Then $\kappa(p)<0$ if and only if there exists an asymptotic coordinate system at $p$(\cite{Sp3}). In this system
$$\kappa(q)=-\frac{\Pi^2(\pl x_1,\pl x_2)}{\det G},\quad\det G=|\pl x_1|^2|\pl x_2|^2-\<\pl x_1,\pl x_2\>^2.$$

In  an asymptotic coordinate system, equation (\ref{3.1}) takes a normal form. We have the following.

\begin{pro}\label{p4.1} Let $M$ be  a hyperbolic orientated surface and let  $\psi(p)=(x_1,x_2):$ $U(\subset M)\rw\R^2$ be an asymptotic coordinate system  on $M$ with the positive orientation.   Then
\be\<D^2w,Q^*\Pi\>=\pm2\sqrt{\frac{-\kappa}{\det G}}w_{x_1x_2}(x)+\mbox{the first order terms},\label{3.5}\ee
where $w(x)=w\circ\psi^{-1}(x)$ and the sign takes $-$ if $\Pi(\pl x_1,\pl x_2)>0$ and $+$ if $\Pi(\pl x_1,\pl x_2)<0,$ respectively.
\end{pro}

{\bf Proof}\,\,\,
Let $p\in U$ be fixed. Let $\a_i=(\a_{i1},\a_{i2})^T\in R^2$ be such that
$$(\a_1,\a_2)\in \SO(2),\quad \det(\a_1,\a_2)=1,\quad G(p)\a_i=\eta_i\a_i\qfq i=1,\,\,2,$$ where $\eta_i>0$ are the eigenvalues of the matrix $G(p).$ Set
\be E_i=\a_{i1}\pl x_1+\a_{i2}\pl x_2\qfq i=1,\,\,2.\label{x2}\ee Since
$$\<E_i,E_j\>=\a_i^TG(p)\a_j=\eta_j\delta_{ij}\qfq 1\leq i,\,\,j\leq2,$$ $\dfrac{E_1}{\sqrt{\eta_1}},$ $\dfrac{E_2}{\sqrt{\eta_2}}$ forms an orthonormal basis of $M_p.$ Moreover, $\frac{E_1}{\sqrt{\eta_1}},$ $\frac{E_2}{\sqrt{\eta_2}}$ is of the positive orientation due to
$$\det\Big(\frac{E_1}{\sqrt{\eta_1}},\frac{E_2}{\sqrt{\eta_2}},\n\Big)=\det[\Big(\pl x_1,\pl x_2,\n\Big)\left(\begin{array}{ccc}\frac{\a_{11}}{\sqrt{\eta_1}}&\frac{\a_{21}}{\sqrt{\eta_2}}&0\\
\frac{\a_{12}}{\sqrt{\eta_1}}&\frac{\a_{22}}{\sqrt{\eta_2}}&0\\
0&0&1\end{array}\right)]=\det\Big(\pl x_1,\pl x_2,\n\Big)\frac{1}{\sqrt{\eta_1\eta_2}}=1.$$ It follows from (\ref{1.3n}) that
$$Q\frac{E_1}{\sqrt{\eta_1}}=-\frac{E_2}{\sqrt{\eta_2}},\quad Q\frac{E_2}{\sqrt{\eta_2}}=\frac{E_1}{\sqrt{\eta_1}}.$$
Using the above relations and the formulas (\ref{3.4}), we have at $p$
\beq \eta_1\eta_2\<D^2w,Q^*\Pi\>&&=D^2w(E_1,E_1)\Pi(E_2,E_2)-2D^2w(E_1,E_2)\Pi(E_1,E_2)\nonumber\\
&&\quad+D^2w(E_2,E_2)\Pi(E_1,E_1)\nonumber\\
&&=2[\a_{21}\a_{22}D^2w(E_1,E_1)-(\a_{11}\a_{22}+\a_{12}\a_{21})D^2w(E_1,E_2)\nonumber\\
&&\quad+\a_{11}\a_{12}D^2w(E_2,E_2)]\Pi(\pl x_1,\pl x_2)\nonumber\\
&&=2D^2w(\a_{21}E_1-\a_{11}E_2,\a_{22}E_1-\a_{12}E_2)\Pi(\pl x_1,\pl x_2)\nonumber\\
&&=-2(\a_{11}\a_{22}-\a_{12}\a_{21})^2D^2w(\pl x_2,\pl x_2)\Pi(\pl x_1,\pl x_2)\nonumber\\
&&=-2[w_{x_1x_2}-D_{\pl x_1}\pl x_2(w)]\Pi(\pl x_1,\pl x_2),\label{3.6}\eeq where the formula
$$\a_{11}\a_{22}-\a_{12}\a_{21}=\det(\a_1,\a_2)=1,$$ has been used.

(\ref{3.5}) follows from (\ref{3.6}) since $\kappa=-\dfrac{\Pi^2(\pl x_1,\pl x_2)}{\eta_1\eta_2}.$ \hfill$\Box$

\begin{lem}\label{l4.1}
There is a $\si_0>0$ such that, for all $p\in \overline{\Om},$ there exist asymptotic coordinate systems $\psi:$ $B(p,\si_0)\rw\R^2$ with
$\psi(p)=(0,0),$
where $B(p,\si_0)$ is the geodesic plate in $M$ centered at $p$ with radius $\si_0.$
\end{lem}

{\bf Proof.}\,\,\,For $p\in\overline{\Om},$ let $\si(p)$ denote the least upper bound of the radii $\si$ for which an asymptotic systems $\psi=x:$ $B(p,\si)\rw\R^2$ with $\psi(p)=(0,0)$
 exists. From the existence of local asymptotic coordinate systems, $\si(p)>0$ for all $p\in\overline{\Om}.$ Let $p,$ $q\in \overline{\Om},$ and $q\in B(p,\si(p)).$
Let
$$\si_1(q)=\inf_{z\in M,\,d(z,p)=\si(p)}d(q,z),$$
where $d(\cdot,\cdot)$ is the distance function on $M\times M$ in the induced metric. Then $\si_1(q)>0$ and $B(q,\si_1(q))\subset B(p,\si(p)),$ since $q\in B(p,\si(p)).$

For any $0<\hat\si<\si_1(q),$ $\overline{B(q,\hat\si)}\subset B(p,\si(p)).$ Thus, there is a $0<\si<\si(p)$ such that $\overline{B(q,\hat\si)}\subset B(p,\si).$ Let $\psi=x:$ $B(p,\si)\rw\R^2$ be
an asymptotic system with $\psi(p)=(0,0).$
Set $\hat\psi(z)=\psi(z)-\psi(q)$ for $z\in B(q,\hat\si).$ Then $\hat\psi:$ $B(q,\hat\si)\rw \R^2$ is an  asymptotic coordinate
system with $\hat\psi(q)=(0,0),$ that is,
$$\si(q)\geq\si_1(q)\qfq q\in B(p,\si(p)).$$   Thus, $\si(p)$ is lower semi-continuous
in $\overline{\Om}$ and $\min_{p\in\overline{\Om}}\si(p)>0$ since $\overline{\Om}$ is compact. \hfill$\Box$

\begin{lem}\label{l4.2} Let $\gamma:$ $[0,a]\rw M$ be a regular curve without self intersection points. Then there is a $\si_0>0$ such that, for all $p\in\{\,\gamma(t)\,|\,t\in(0,a)\,\},$ $S(p,\si_0)$ has at most two
intersection points with $\{\,\gamma(t)\,|\,t\in[0,a]\,\},$ where $S(p,\si_0)$ is the geodesic circle centered at $p$ with radius $\si_0.$ If $p=\gamma(0),$ or $\gamma(a),$ then  $S(p,\si_0)$
has at most
one intersection point with $\{\,\gamma(t)\,|\,t\in[0,a]\,\}.$
\end{lem}

{\bf Proof.}\,\,\,By contradiction. Let the claim in the lemma be not true. For each integer $k\geq1,$ there exists $t_k<t_k^1<t_k^2$(or $t_k>t_k^1>t_k^2$) in $[0,a]$ such that
\be d(\gamma(t_k),\gamma(t_k^1))=d(\gamma(t_k),\gamma(t_k^2))=\frac{1}{k}\qfq k\geq 1.\label{x3.20}\ee

We may assume that
$$t_k\rw t^0,\quad t_k^1\rw t^1,\quad t_k^2\rw t^2\quad \mbox{as}\quad k\rw \infty, $$ for certain points $t^0,$ $t^1,$ $t^2\in[0,a].$ Then $0\leq t^0\leq t^1\leq t^2\leq a$ and
$$\gamma(t^0)=\gamma(t^1)=\gamma(t^2).$$ The assumption that the curve $\gamma$ has no self intersection point implies that
$$t^0=t^1=t^2.$$

For $k\geq1, $ let
$$f_k(t)=\frac{1}{2}\rho_k^2(\gamma(t)), \qfq t\in[0,a],$$ where $\rho_k(p)=d(\gamma(t_k),p)$ for $p\in M.$
It follows from (\ref{x3.20}) that there is a $\zeta_k$ with $t_k^1<\zeta_k<t_k^2$ such that
$$f_k'(\zeta_k)=0.$$ On the other hand, the formula $f'_k(t)=\rho_k(\gamma(t))\<D\rho_k(\gamma(t)),\dot\gamma(t)\>$ implies that $f'_k(t_k)=0.$ Thus, we obtain $\eta_k\in(t_k,\zeta_k)$ such that
$$f''_k(\eta_k)=0\qfq k\geq1.$$ Since
$$f''_k(t)=D(\rho_kD\rho_k)(\dot\gamma(t),\dot\gamma(t))+\rho_k(\gamma(t))\<D\rho_k(\gamma(t)),D_{\dot\gamma(t)}\dot\gamma\>,$$ we have
$$|\dot\gamma(t^0)|^2=f_0''(t^0)=\lim_{k\rw\infty}f_k''(\eta_k)=0,$$ which contradicts the regularity of the curve $\gamma,$ where
$$f_0(t)=\frac{1}{2}d^2(\gamma(t^0),\gamma(t))\qfq t\in[0,a].$$ \hfill$\Box$

We need the following.
\begin{pro}\label{p10} $(i)$\,\,\,
$\det\Big(Q\na\n X,X,\n(x)\Big)=\Pi(X,X)(x)$ for $X\in T_xM,$ $x\in M.$

$(ii)$\,\,\,$\Pi(Q\nabla\n X,Q\nabla\n X)=\kappa\Pi(X,X)$ for $X\in T_xM,$ $x\in M.$
\end{pro}

{\bf Proof}\,\,\,Let $x\in M$ be given. Let $e_1,$ $e_2$ be an orthonormal basis of $T_xM$ with the positive orientation such that
\be\Pi(e_i,e_j)(x)=\lam_i\delta_{ij}\qfq1\leq i,\,j\leq2.\label{n4.3x}\ee Then
\beq\det\Big(Q\nabla\n X, X,\n\Big)=\det\Big(e_1,e_2,\n\Big)\left(\begin{array}{ccc}\lam_2\<X,e_2\>&\<X,e_1\>&0\\
-\lam_1\<X,e_1\>&\<X,e_2\>&0\\
0&0&1\end{array}\right)=\Pi(X,X).\eeq In addition,  using (\ref{n4.3x}), we have
\beq \Pi(Q\nabla\n X,Q\nabla\n X)&&=\Pi\Big(-\lam_1\<X,e_1\>e_2+\lam_2\<X,e_2\>e_1,\,-\lam_1\<X,e_1\>e_2+\lam_2\<X,e_2\>e_1\Big)\nonumber\\
&&=\lam_1^2\lam_2\<X,e_1\>^2+\lam_2^2\lam_1\<X,e_2\>^2=\kappa\Pi(X,X).\nonumber\eeq   \hfill$\Box$

\begin{lem}\label{l4.3} Let $p_0\in M$ and let $B(p_0,\si)$ be the geodesic ball centered at $p_0$ with radius $\si>0.$ Let $\gamma:$ $[-a,a]\rw B(p_0,\si)$ and $\b:$ $[-b,b]\rw B(p_0,\si)$ be two noncharacteristic curves of class $\CC^1,$ respectively, with
$$\gamma(0)=\b(0)=p_0,\quad \Pi(\dot\gamma(0),\dot\b(0))=0.$$
Let $\hat\psi:$ $B(p_0,\si)\rw\R^2$ be an  asymptotic coordinate system. Then there exists an  asymptotic coordinate system  $\psi:$ $B(p_0,\si)\rw\R^2$
with $\psi(p_0)=(0,0)$  such that
\be\psi(\gamma(t))=(t,-t)\qfq t\in[-a,a],\label{ch4.27}\ee
\be\b_1'(s)>0,\quad \b_2'(s)>0\qfq s\in[-b,b],\label{4..16}\ee where
$\psi(\b(s))=(\b_1(s),\b_2(s)).$ Moreover, for $X=X_1\pl x_1+X_2\pl x_2$ with $\Pi(X,X)\not=0,$ we have
\be\varrho(X)Q\na\n X=\chi\Big(\g'(0),\b'(0)\Big)\left\{\begin{array}{l} X_1\pl x_1-X_2\pl x_2,\quad X_1X_2>0,\\
-X_1\pl x_1+X_2\pl x_2,\quad X_1X_2<0,\end{array}\right.\label{x4.24}\ee where  $\varrho(X)$ is given in $(\ref{rho4.3})$ and
$$\chi\Big(\g'(0),\b'(0)\Big)=\sign\det\Big(\g'(0),\b'(0),\n(p_0)\Big).$$
\end{lem}

{\bf Proof.}\,\,\,Let $\hat\psi(p_0)=(0,0)$ and $$\hat\psi(\gamma(t))=(\gamma_1(t),\gamma_2(t))\qfq t\in[-a,a].$$ Since $\gamma$ is noncharacteristic,
$$\Pi(\dot\gamma(t),\dot\gamma(t))=2\gamma_1'(t)\gamma_2'(t)\Pi(\pl x_1,\pl x_2)\not=0\qfq t\in[-a,a].$$
Without loss of generality, we assume that
\be\gamma_1'(t)>0,\quad\gamma_2'(t)<0\qfq t\in[-a,a].\label{xx4.16}\ee
We extend the domain $[-a,a]$ of $\gamma(t)$ to $\R$ such that
$$\lim_{t\rw\pm\infty}\gamma_1(t)=\pm\infty,\quad \lim_{t\rw\pm\infty}\gamma_2(t)=\mp\infty,$$
and the relations (\ref{xx4.16}) hold for all $t\in\R.$ Consider a diffeomorphism $\var(x)=y:$ $R^2\rw\R^2$ given by
\be\var(x)=(\gamma_1^{-1}(x_1),-\gamma_2^{-1}(x_2))\qfq x=(x_1,x_2)\in\R^2.\label{4..18}\ee
Then $\var\circ\hat\psi:$ $B(p_0,\si)\rw\R^2$ is an  asymptotic coordinate system such that
\be\var\circ\hat\psi(\gamma(t))=(t,-t)\qfq t\in[-a,a].\label{4..17}\ee

Let $\var\circ\hat\psi(\b(s))=(\b_1(s),\b_2(s)).$ Since $\b$ is noncharacteristic,
$$\b_1'(s)\b_2'(s)\not=0\qfq s\in[-b,b].$$ In addition, the assumption
$\Pi(\dot\gamma(0),\dot\b(0))=0$ and the relation (\ref{4..17}) imply that
$$0=\Pi(\dot\gamma(0),\dot\b(0))=\Pi(\pl x_1-\pl x_2,\,\b_1'(0)\pl x_1+\b_2'(0)\pl x_2)=[\b_2'(0)-\b_1'(0)]\Pi(\pl x_1,\pl x_2),$$ that is,
$\b_1'(0)=\b_2'(0).$ If $\b_1'(0)>0,$ we let $\psi(p)=\var\circ\hat\psi(p)$ to have (\ref{4..16}).
If $\b_1'(0)<0,$ we define instead of (\ref{4..18})
$$\var(x)=(\gamma_2^{-1}(x_2),-\gamma_1^{-1}(x_1))\qfq x=(x_1,x_2)\in\R^2.$$
Thus (\ref{4..16}) follows again.

Next, we prove (\ref{x4.24}). Let $Q\na\n X=Y_1\pl x_1+Y_2\pl x_2.$
Since  $(Y_1X_2+Y_2X_1)\Pi(\pl x_1,\pl x_2)=\Pi\Big(Q\nabla\n X,X\Big)=\<Q\nabla\n X,\nabla\n X\>=0,$
we have
$$Q\nabla\n X=\si(X_1\pl x_1-X_2\pl x_2),$$ where $\si$ is a function.
Using Proposition \ref{p10} (ii), we obtain
$$\si^2=-\kappa.$$

Next, from (\ref{ch4.27}) and (\ref{4..16}), we have
$$\Big(\g'(0),\b'(0),\n\Big)=\Big(\pl x_1,\pl x_2,\n\Big)\left(\begin{array}{ccc}1&\b_1'(0)&0\\
-1&\b_2'(0)&0\\
0&0&1\end{array}\right), $$ which yields
$$\sign\det \Big(\g'(0),\b'(0),\n\Big)=\sign\det\Big(\pl x_1,\pl x_2,\n\Big).$$
Thus (\ref{x4.24}) follows from Proposition \ref{p10} (i).
\hfill$\Box$

Denote
\be\Om(0,s_0)=\{\,\a(t,s)|\,t\in(0,a),\,\,s\in(0,s_0)\,\}\qfq s_0\in[0,b].\label{gg4.30}\ee Then $\Om=\Om(0,b).$

\begin{lem}\label{l4.4} Let the assumptions in Theorem $\ref{t4.1}$ hold. Then
there is a $0<\omega\leq b$ such that problem $(\ref{3.1})$ admits a unique
solution $w\in \CC^{m,1}(\overline{\Om(0,\omega)})$  with the  data
$(\ref{4.3})$  where $s\in[0,\omega],$ and $(\ref{x1})$  to satisfy
\beq\|w\|_{\CC^{m,1}(\overline{\Om(0,\omega)})}&&\leq C(\|p_1\|_{\CC^{m-1,1}[0,b]}+\|p_2\|_{\CC^{m-1,1}[0,b]}+\|q_0\|_{\CC^{m,1}[0,a]}\nonumber\\
&&\quad+\|q_1\|_{\CC^{m-1,1}[0,a]}
+\|f\|_{\CC^{m-1,1}(\overline{\Om})}).\label{xn4.24} \eeq
\end{lem}

{\bf Proof.}\,\,\,Let $\si_0>0$ be given small such that the claims in Lemmas \ref{l4.1} and \ref{l4.2} hold, where $\gamma(t)=\a(t,0)$ in Lemma \ref{l4.2}.
We divide the curve $\a(t,0)$ into $m$ parts with the points $\lam_i=\a(t_i,0)$  such that
$$\lam_0=\a(0,0),\quad \lam_m=\a(a,0),\quad d(\lam_i,\lam_{i+1})=\frac{\si_0}{3},\quad 0\leq i\leq m-2,\quad d(\lam_{m-1},\lam_m)\leq\frac{\si_0}{3},$$
where $t_0=0,$ $t_1>0,$ $t_2>t_1,$ $\cdots,$  and $t_m=a>t_{m-1}.$ For simplicity, we assume that $m=3.$ The other cases can be treated by a similar argument.

We shall construct a local solution in a neighborhood of $\a(t,0)$ by the following steps.

{\bf Step 1.}\,\,\,Let $\hat s_0>0$ be small such that
$$\a(0,s)\in B(\lam_0,\si_0)\qfq s\in[0,s_0].$$
From Lemma \ref{l4.3}, there is  asymptotic coordinate system $\psi_0(p)=x:$ $B(\lam_0,\si_0)\rw\R^2$  with $\psi_0(\lam_0)=(0,0)$ such that
\be\psi_0(\a(t,0))=(t,-t)\qfq t\in[0,t_2],\label{4.16}\ee
$$\b_{01}'(s)>0,\quad \b_{02}'(s)>0\quad \mbox{for all}\quad s\in[0,s_0],$$ where $\b_0(s)=\psi_0(\a(0,s))=(\b_{01}(s),\b_{02}(s)).$  Let $\g_0(t)=(t,-t).$ We may assume that $s_0>0$ is given small such that $\b_{01}(s_0)\leq t_2$ since $\b_{01}(0)=0.$ Set
\beq\Xi_1(\b_0,\g_0)&&=P_1(\b_0)\cup R((\b_{01}(s_0),0),c_0,d_0)\cup E(\g_0),\label{nx4.34}\eeq as in (\ref{set3.15}) with $c_0=t_2-\b_{01}(s_0)$ and
$d_0=\b_{02}(s_0).$ Then we let
$$\Om_0=\Om\cap\psi_0^{-1}[\Xi_1(\b_0,\g_0)].$$

Noting that for the region $\Om_0$
$$\chi(\mu(\a(t,0)),\a_t(t,0))=\chi(-\a_s(0,0),\a_t(0,0))\qfq t\in(0,t_2),$$
$$\chi(\mu(\a(0,s)),\a_s(0,s))=\chi(-\a_t(0,0),\a_s(0,0))\qfq s\in(0,s_0),$$
from (\ref{x4.24}), we obtain
\be \T_1\a_s(0,s)=\b_{01}'(s)\pl x_1,\quad \T_2\a_s(0,s)=\b_{02}'(s)\pl x_2\qfq s\in(0,s_0),\label{Tt1}\ee
\be \T_1\a_t(t,0)=\pl x_1,\quad \T_2\a_t(t,0)=-\pl x_2\qfq t\in(0,t_2). \label{Tt2}\ee

From Proposition \ref{p4.1}, solvability of problem (\ref{3.1}) on $\Om\cap \psi_0^{-1}(\Xi_1(\b_0,\g_0))$ is equivalent to
that of problem (\ref{3.7}) over the region $\Xi_1(\b_0,\g_0).$ Next, we consider the transfer of the boundary data under the chart $\psi_0.$ The corresponding part data are
$$w_{x_2}\circ\b_{0}(s)=\<Dw,\T_2\a_s\>\circ\a(0,s)/\b_{02}'(s)=p_1(s)/\b_{02}'(s)\qfq s\in[0,s_0],$$
$$w(t,-t)=w\circ\psi_0^{-1}(t,-t)=w(\a(t,0))=q_0(t)\qfq t\in[0,t_2],$$
$$\frac{\pl}{\pl\nu}w(t,-t)=\frac{1}{\sqrt{2}}\<Dw,(\T_2-\T_1)\a_t\>\circ\a(t,0)=q_1(t)\qfq t\in[0,a],$$ where
$$w(x)=w\circ\psi^{-1}(x).$$
It is easy to check
that $p_1/\b_{02}',$  $q_0,$ $q_1,$ and $f$ are  $m$th order compatible at $\a(0,0)$ in the sense of Definition \ref{d4.2} is equivalent to that $p_1/\b_{02}',$  $q_0,$ $q_1,$ and $\hat f$ do
in the sense of Definition \ref{d3.1}, where
$$\hat f=\frac{f\circ\psi_0^{-1}(x)}{2}\sqrt{\frac{\det G(x)}{-k\circ\psi_0^{-1}(x)}}\qfq x\in\psi_0(B(\lam_0,\si_0))\subset\R^2, $$ where
$\det G(x)=\det(\<\pl x_i,\pl x_j\>).$

From Proposition \ref{p3.10}, problem (\ref{3.7}) admits a unique solution $w\in \CC^{m,1}(\overline{\Xi_1(\b_0,\g_0}))$ with the corresponding boundary data.
 Thus, we have obtained a solution, denoted by
$w_0\in \CC^{m,1}(\overline{\Om_0}),$ to problem (\ref{3.1}) with the data
$$\<Dw_0,\T_2\a_s\>\circ\a(0,s)=p_1(s)\qfq s\in[0,s_0],$$
$$w_0\circ\a(t,0)=q_0(t),\quad\frac{1}{\sqrt{2}}\<Dw_0,(\T_2-\T_1)\a_t\>\circ\a(t,0)=q_1(t)\qfq t\in[0,t_2],$$ where
$$\Om_0=\Om\cap\psi_0^{-1}[\Xi_1(\b_0,\g_0)].$$
It follows from the estimate in Proposition \ref{p3.10} that
\beq\|w_0\|_{\CC^{m,1}(\overline{\Om_0})}\leq C\Ga_{m\CC}(p_1,p_2,q_0,q_1,f),\label{xn4.28}\eeq where
\beq\Ga_{m\CC}(p_1,p_2,q_0,q_1,f)&&=\|p_1\|_{\CC^{m-1,1}(0,b)}+\|p_2\|_{\CC^{m-1,1}(0,b)}+\|q_0\|_{\CC^{m,1}(0,a)}\nonumber\\
&&\quad+\|q_1\|_{\CC^{m-1,1}(0,a)}+\|f\|_{\CC^{m-1,1}(\overline{\Om})}.\nonumber\eeq

We define a curve on $\Om_0$ by
\be\zeta_1(s)=\psi_0^{-1}\circ\gamma_{t_1}(s)\qfq s\in[0,s_{t_1}],\label{b4.32}\ee where
$$\gamma_{t_1}(s)=(s+t_1,s-t_1),\quad s_{t_1}=\left\{\begin{array}{l}t_1\quad\mbox{if}\quad t_1\in(0,\dfrac{t_2}{2}],\\
t_2-t_1\quad\mbox{if}\quad t_1\in(\dfrac{t_2}{2},t_2).\end{array}\right. $$
Then $\zeta_1(s)$ is noncharacteristic and
\be\Pi(\dot\zeta_1(0),\a_t(t_1,0))=\Pi(\pl x_1+\pl x_2,\pl x_1-\pl x_2)=0.\label{xx4.17}  \ee

{\bf Step 2.}\,\,\, Let the curve $\zeta_1$ be given in (\ref{b4.32}). Let $s_1>0$ be small such that
$$\zeta_1(s)\in B(\lam_1,\si_0)\qfq s\in[0, s_1].$$
From the noncharacteristicness of $\zeta_1(s)$ and the relation (\ref{xx4.17}) and Lemma \ref{l4.3} again, there exists  an  asymptotic coordinate system  $\psi_1(p)=x:$ $B(\lam_1,\si_0)\rw\R^2$ with $\psi_1(\lam_1)=(0,0)$ and
$$\psi_1(\a(t+t_1,0))=(t,-t)\qfq t\in[0,t_3-t_1],$$
$$\b_{11}'(s)>0,\quad\b_{12}'(s)>0\qfq s\in[0,s_1],$$ where $\b_1(s)=\psi_1(\zeta_1(s))=(\b_{11}(s),\b_{12}(s)).$ We also assume that $s_1>0$ has been taken small such that $\b_{11}(s_1)\leq t_3.$
This time, we set
$$\Xi_1(\b_1,\g_1)=P_1(\b_1)\cup R((\b_{11}(s_1),0),c_1,d_1)\cup E_1(\g_1),$$ where $c_1=t_3-t_1-\b_{11}(s_1)$, $d_1=\b_{12}(s_0),$
and $\g_1(t)=\psi_1(\a(t+t_1,0)).$ Next, let
$$\Om_1=\Om\cap\psi_1^{-1}[\Xi_1(\b_1,\g_1)].$$
Since for the region $\Om_1$
$$\chi(\mu(\zeta_1(s)),\zeta_1'(s))=\chi(-\a_t(0,t_1),\zeta_1'(0))\qfq s\in(0,s_1),$$
$$\chi(\mu(\a(t_1+t,0)),\a_t(t_1+t,0))=\chi(-\zeta_1'(0),\a_t(t_1,0))\qfq t\in(0,t_3-t_1),$$ it follows from (\ref{x4.24}) that
$$ \T_1\zeta_1'(s)=\b_{11}'(s)\pl x_1,\quad \T_2\zeta_1'(s)=\b_{12}'(s)\pl x_2\qfq s\in(0,s_1),$$
$$ \T_1\a_t(t_1+t,0)=\pl x_1,\quad \T_2\a_t(t_1+t,0)=-\pl x_2\qfq t\in(0,t_3-t_1). $$

 By some similar arguments in Step 1, we obtain a unique solution
$w_1\in \CC^{m,1}(\overline{\Om_1})$ to problem (\ref{3.1}) with the data
$$\<Dw_1,\T_2\dot\zeta_1\>\circ\b_1(s)=\<Dw_{0},\T_2\dot\zeta_1(s)\>\circ\b_1(s)\qfq s\in[0,s_1],$$
$$w_1(\a(t,0))=q_0(t),\quad \frac{1}{\sqrt{2}}\<Dw_1,(\T_2-\T_1)\a_t\>\circ\a(t,0)=q_1(t)\qfq t\in[t_1,t_3],$$ 
where $w_0$ is the solution of (\ref{3.1}) on $\Om_0,$ given in Step 1. The following estimate also holds
\beq\|w_1\|_{\CC^{m,1}(\overline{\Om_1})}&&\leq C(\|\<Dw_{0},\T_2\dot\zeta_1\>\circ\zeta_1\|_{\CC^{m-1,1}[0,s_1]}+\|q_0\|_{\CC^{m,1}[0,a]}+\|q_1\|_{\CC^{m-1,1}[0,a]}\nonumber\\
&&\quad+\|f\|_{\CC^{m-1,1}(\overline{\Om})})\leq C\Ga_{m\CC}(p,q_0,q_1,h).\label{n4.31}\eeq

As in Step 1, we define a curve on $\Om_1$ by
\be\zeta_2(s)=\psi_1^{-1}(s+t_2-t_1,s+t_1-t_2)\qfq s\in[0,s_{t_2}],\label{zeta}\ee where
$$  s_{t_2}=t_2-t_1\quad\mbox{if}\quad t_2-t_1\leq\frac{t_3-t_1}2;\quad  s_{t_2}=t_3-t_2\quad\mbox{if}\quad t_2-t_1>\frac{t_3-t_1}2.$$
 Then $\zeta_2(s)$ is noncharacteristic and
\be\Pi(\dot \zeta_2(0),\a_t(t_2,0))=\Pi(\pl x_1+\pl x_2,\pl x_1-\pl x_2)=0.\label{x4.17}  \ee

{\bf Step 3.}\,\,\, Let the curve $\zeta_2$ be given in (\ref{zeta}). Let $s_2>0$ be small such that
$$\zeta_2(s),\quad\a(a,s)\in B(\lam_2,\si_0)\qfq s\in[0,s_2].$$
Let $\psi_2(p)=x:$ $B(\lam_2,\si_0)\rw\R^2$ be an  asymptotic coordinate system with $\psi_2(\lam_2)=(0,0),$
\be\psi_2(\a(t+t_2,0))=(t,-t)\qfq t\in[0,a-t_2],\label{4.27}\ee and
$$\b_{21}'(s)>0,\quad \b_{22}'(s)>0\qfq s\in[0, s_2],$$ where $\b_2(s)=\psi_2(\zeta_2(s))=(\b_{21}(s),\b_{22}(s)).$

Let $\b_3(s)=\psi_2(\a(a,s))=(\b_{31}(s),\b_{32}(s)).$
Next, we prove that
\be\b_{31}'(s)>0,\quad \b_{32}'(s)>0\qfq s\in[0, s_2],\label{4.28}\ee by contradiction.
Since $\a(a,s)$ is noncharacteristic, using (\ref{4.27}) and the assumption $\Pi(\a_t(a,0),\a_s(a,0))=0,$ we have
$$\b_{31}'(0)=\b_{32}'(0);\quad\mbox{thus}\quad\b_{31}'(s)\b_{32}'(s)>0\qfq s\in[0, s_2].$$ Let
$$p(t,s)=\a_1(t,s)+\a_2(t,s),\quad\psi_2(\a(t+t_2,s))=(\a_1(t,s),\a_2(t,s)).$$
 Let (\ref{4.28}) be not true, that is,
$\b_{31}'(s)<0,$  $\b_{32}'(s)<0\qfq s\in[0, s_2].$ Thus
$$p(0,s)=\b_{21}(s)+\b_{22}(s)> \b_{21}(0)+\b_{22}(0)=0\qfq s\in(0,s_2],$$
$$p(a-t_2,s)=\b_{31}(s)+\b_{32}(s)< \b_{31}(0)+\b_{32}(0)=0\qfq s\in(0, s_2].$$ Let $t(s)\in(0,a-t_2)$ be such that
\be\a_1(t(s),s)+\a_2(t(s),s)=0\qfq s\in(0, s_2).\label{4.29}\ee Since $\a_{1t}(0,0)=1$ and $\a(t+t_2,s)$ are noncharacteristic for all $s\in[0,s_2],$ we have $\a_{1t}(t,s)>0$ and
$$0<\a_1(0,s)<\a_1(t(s),s)<\a_1(a-t_2,s)=\b_{31}(s)<\b_{31}(0)=a-t_2.$$ Thus, equality (\ref{4.29}) means that $\a(\a_1(t(s),s)+t_2,0)=\a(t(s),s),$ which is a contradiction since
 $\a:$ $[0,a]\times[a,b]\rw M$ is an imbedding map.

We also assume that $s_2$ has been taken so small such that
$$\b_{21}(s_2)<a-t_2,\quad \b_{32}(s_2)<0,$$ since $\b_{21}(0)=0$ and $\b_{32}(0)=-(a-t_2)<0.$  Let $\g_2(t)=\psi_2(\a(t+t_2,0))=(t,-t).$ We now set
$$\Phi(\b_2,\g_2,\b_3)=\Xi_1(\b_2,\g_2)\cup R((a-t_2,\b_{32}(s_2)),c_3,d_3)\cup P_2(\b_3),$$ where $\Xi_1(\b_2,\g_2),$ $R((a-t_2,\b_{32}(s_2)),c_3,d_3),$
and $P_2(\b_3)$ are given in (\ref{set3.15}), (\ref{R}), and (\ref{P_2}), respectively, with $c_3=\b_{32}(s_2)-a+t_2$ and $d_3=\b_{22}(s_2)-\b_{32}(s_2).$ Let
$$\Om_2=\Om\cap\psi_2^{-1}[\Phi(\b_2,\g_2,\b_3)].$$

This time we use (\ref{x4.24}) to obtain, for the region $\Om_2,$
$$ \T_1\zeta_2'(s)=\b_{21}'(s)\pl x_1,\quad \T_2\zeta_2'(s)=\b_{22}'(s)\pl x_2\qfq s\in(0,s_2),$$
$$ \T_1\a_t(t_2+t,0)=\pl x_1,\quad \T_2\a_t(t_2+t,0)=-\pl x_2\qfq t\in(0,a-t_2),$$
$$ \T_1\a_s(a,s)=\b_{32}'(s)\pl x_2,\quad \T_2\a_s(a,s)=\b_{31}'(s)\pl x_1\qfq s\in(0,s_2).$$

Applying Proposition \ref{t3.1}, problem (\ref{3.1}) admits a unique solution $w_2\in \CC^{m,1}(\overline{\Om}_2)$ with the data
$$\<Dw_2,\T_2\dot\zeta_2\>\circ\b_2(s)=\<Dw_1,\T_2\dot\zeta_2\>\circ\b_2(s),\quad \<Dw_2,\T_2\a_s\>\circ\a(a,s)=p_2(s)\qfq s\in[0,s_2],$$
$$w_2(\a(t,0))=q_0(t),\quad \frac{1}{\sqrt{2}}\<Dw_2,(\T_2-\T_1)\a_t\>\circ\a(t,0)=q_1(t)\qfq t\in[t_2,a].$$ 
Using the estimates in Proposition \ref{t3.1} and (\ref{n4.31}), we obtain
\beq\|w_2\|_{\CC^{m,1}(\overline{\hat\Om_2})}&&\leq C(\|\<Dw_{1},\T_2\dot\zeta_2\>\circ\zeta_2\|_{\CC^{m-1,1}[0,s_2]}+\|p_2\|_{\CC^{m-1,1}[0,b]}+\|q_0\|_{\CC^{m,1}[0,a]}\nonumber\\
&&\quad+\|q_1\|_{\CC^{m-1,1}[0,a]}+\|f\|_{\CC^{m-1,1}(\overline{\Om})})\leq C\Ga_{m\CC}(p_1,p_2,q_0,q_1,f).\label{xn4.36}\eeq

{\bf Step 4.}\,\,\,We define
$$w=w_i\qfq p\in\Om_i\qfq i=0,\,1,\,2.$$ Let $\omega>0$ be small such that
$$\a(t,s)\in\Om_0\cup\Om_1\cup\Om_2\qfq (t,s)\in(0,a)\times(0,\omega).$$
Then $w\in \CC^{m,1}(\overline{\Om(0,\omega)})$ will be a solution to (\ref{3.1}) with the corresponding data if we show that
\be w_0(p)=w_1(p)\qfq p\in \Om_0\cap\Om_1;\quad w_1(p)=w_2(p)\qfq p\in \Om_1\cap\Om_2.\label{xx4.19}\ee
Since
$$w_{1x_2}\circ\b_1(s)=w_{0x_2}\circ\b_1(s)\qfq s\in[0,s_1],$$
$$w_0(t,-t)=w_1(t,-t),\quad \frac{\pl w_0}{\pl\nu}(t,-t)=\frac{\pl w_1}{\pl\nu}(t,-t)\qfq t\in[t_1,t_2],$$
from the uniqueness in Proposition \ref{p3.10}, we have
$$w_0(x)=w_1(x)\qfq x\in \Xi_1(\b_0,\g_0)\cap\Xi_1(\b_1,\g_1),$$ which yields the first identity in (\ref{xx4.19}).
A similar argument shows that the second identity in (\ref{xx4.19}) is true.

Finally, the estimate (\ref{xn4.24}) follows  from  (\ref{xn4.28}), (\ref{n4.31}), and (\ref{xn4.36}).
\hfill$\Box$

From a similar argument as for the proof of Lemma \ref{l4.4}, we obtain the following.

\begin{lem}\label{l4.5} Let the assumptions in Theorem $\ref{t4.2}$ hold. Then
there is a $0<\omega\leq b$ such that problem $(\ref{3.1})$ admits a unique
solution $w\in\WW^{2,2}(\Om(0,\omega))$  with the  data
$(\ref{4.3})$  where $s\in(0,\omega),$ and $(\ref{x1})$  to satisfy
\beq\|w\|^2_{\WW^{2,2}(\Om(0,\omega))}&&\leq C(\|q_0\|^2_{\WW^{2,2}(0,a)}+\|q_1\|^2_{\WW^{1,2}(0,a)}+\|p_1\|^2_{\WW^{2,2}(0,b)}\nonumber\\
&&\quad+\|p_2\|^2_{\WW^{2,2}(0,b)}+\|f\|_{\WW^{1,2}(\Om)}).\label{x4.43}\eeq
\end{lem}

\vskip2mm

We are now ready to prove Theorems \ref{t4.1} and \ref{t4.2}.

{\bf Proof of Theorem \ref{t4.1}}\,\,\,Let $\aleph$ be the set of all $0<\omega\leq b$ such that the claims in Lemma \ref{l4.4} hold. We shall prove
$$b\in \aleph.$$

Let $\omega_0=\sup_{\omega\in\aleph}\omega.$  Then $0<\omega_0\leq b.$
Thus there is a unique solution $w\in \CC^{m,1}(\Om(0,\omega_0))$ to (\ref{3.1})  with the  data
$(\ref{4.3}),$  where $s\in[0,\omega_0),$ and $(\ref{x1}).$

Next we show that $\omega_0=b$ by contradiction. Let $0<\omega_0<b.$  By an argument as for Lemma \ref{l4.4}, the solution $w\in\CC^{m,1}(\Om(0,\omega_0))$ can be extended such
that $w\in\CC^{m,1}(\overline{\Om(0,\omega_0))}.$  Then by Lemma \ref{l4.4} again, $w$ can be extend outside $\CC^{m,1}(\overline{\Om(0,\omega_0)}),$ which contradicts with the definition of $\omega_0.$

 Let $\lam_0=\a(0,\omega_0),$ $\si_0,$ $t_0=0,$ $t_1,$ $t_2,$ and $t_3=a$  be given as in the proof of Lemma \ref{l4.4}.

Let $\psi_0(p)=x:$ $B(\lam_0,\si_0)\rw\R^2$ be an  asymptotic coordinate system with $\psi_0(\lam_0)=(0,0)$ such that
$$\psi_0(\a(t,\omega_0))=(t,-t)\qfq t\in[0,t_2],$$
$$\z_{1}'(s)>0,\quad\z_{2}'(s)>0\qfq s\in[\omega_0-\varepsilon_0,\omega_0],$$
where $\psi_0(\a(0,s))=(\z_{1}(s),\z_{2}(s)).$

 For $\varepsilon>0,$ let
$$\b_{0}(s)=\psi_0(\a(0,s+\omega_0-\varepsilon))=(\b_{01}(s),\b_{02}(s)),\quad \g_0(t)=\psi(\a(t,\omega_0-\varepsilon))=(\g_{01}(t),\g_{02}(t)),$$ for $s\in[0,2\varepsilon],$ where
$\b_{0i}(s)=\z_i(s+\omega_0-\varepsilon)$ for $i=1,$ $2.$
We fixed $\varepsilon>0$  small such that
$$\omega_0+\varepsilon\leq b,\quad\b_{01}(\varepsilon)\leq\g_{01}(t_2),\quad \g_{01}'(t)>0,\quad\g_{02}'(t)<0.$$
Let $\Xi_1(\b_0,\g_0)$ be given as in (\ref{nx4.34}). Clearly,
$$\{\,\a(t,\omega_0)\,|\,t\in[0,t_2]\,\}\subset\psi_0^{-1}( \overline{\Xi_1(\b_0,\g_0)}).$$
From Proposition \ref{p3.10}, we can extend a solution $w$ such that $w$ is $\CC^{k,1}$ on the segment $ \{\,\a(t,\omega_0+\varepsilon)\,|\,t\in[0,t_2]\,\}.$ Repeating Steps 2-4 in the proof of Lemma \ref{l4.4},
the solution $w$ can be extended such that  $w$ is $\CC^{k,1}$ on the segment $ \{\,\a(t,\omega_0+\varepsilon)\,|\,t\in[0,a]\,\},$ which contradicts the definition of $\omega_0.$ The proof is complete. \hfill$\Box$

{\bf Proof of Theorem \ref{t4.2}}\,\,\,A similar argument as in the proof of Theorem \ref{t4.1} completes the proof. \hfill$\Box$

To prove Theorems \ref{t4.3} and \ref{t4.4}, we need the following lemmas.

\begin{lem}\label{l4.6} Let the assumptions in Theorem $\ref{t4.3}$ hold. Then
there are  $0<\omega\leq b$ and $C>0$ such that for all
solutions $w\in\WW^{2,2}(\Om)$  to problem $(\ref{3.1})$
\be\|w\|^2_{\WW^{2,2}(\Om(0,\omega))}\leq C[\|f\|^2_{\WW^{1,2}(\Om)}+\Ga(\Om,w)],\label{x4.44}\ee where $\Om(0,\omega)$ and $\Ga(\Om,w)$ is given in $(\ref{gg4.30})$ and $(\ref{GaOm}),$ respectively.
\end{lem}

{\bf Proof}\,\,\,We keep all the notion in the proof of Lemma \ref{l4.4}. Let $\omega>0$ be given in Step 4. Then
$$w_0(x)=w\circ\psi_0^{-1}(x)$$ is a solution to problem (\ref{3.7}) on the region $\Xi_1(\b_0,\g_0),$ where $\Xi_1(\b_0,\g_0)$ is given in (\ref{nx4.34}) and
$$\b_0(s)=\psi_0(\a(0,s))=(\b_{01}(s),\b_{02}(s))\qfq s\in(0,s_0),\quad \gamma_0(t)=\psi_0(\a(t,0))=(t,-t)$$ for $t\in[0,t_2].$

It follows from (\ref{Tt1}) and (\ref{Tt2}) that
$$\Big||D^2w(\T_1\a_t(t,0),\T_1\a_t(t,0))|-|w_{0x_1x_1}\circ\gamma_0(t)|\Big|\leq C|\nabla w_0\circ\gamma_0(t)|,$$

Similarly, we have
$$\Big||D^2w(\T_2\a_t(t,0),\T_2\a_t(t,0))|-|w_{0x_2x_2}\circ\gamma_0(t)|\Big|\leq C|\nabla w_0\circ\gamma_0(t)|,$$
$$\Big||D^2w(\T_1\a_s(0,s),\T_1\a_s(0,s))|-|w_{0x_1x_1}\circ\b_0(s)|\b_{01}'^2(s)\Big|\leq C|\nabla w_0\circ\b_0(s)|,$$
$$\Big||D^2w(\T_2\a_s(0,s),\T_2\a_s(0,s))|-|w_{0x_2x_2}\circ\b_0(s)|\b_{02}'^2(s)\Big|\leq C|\nabla w_0\circ\b_0(s)|.$$
Using the above relations, we obtain
\be\Ga(\g_0,w_0)+\Ga_2(\b_0,w_0)\leq C\Ga(\Om,w),\label{4.50}\ee where $\Ga(\g_0,w_0)$ and $\Ga_2(\b_0,w_0)$ are  given in (\ref{g3.7}) and (\ref{gam3.27}), respectively.

Applying Proposition \ref{np3.12} to $\Xi_1(\b_0,\g_0)$ and using (\ref{4.50}), we have
$$\|w\|^2_{\WW^{2,2}(\Om_0)}\leq C\|w_0\|^2_{\WW^{2,2}(\Xi_1(\b_0,\g_0))}\leq C(\|f\|^2_{\WW^{1,2}}+\Ga(\Om,w)).$$ Using (\ref{xnnn3.3}) by a similar argument as for the above estimates, we obtain
$$\|w\|^2_{\WW^{2,2}(\Om_i)}\leq C(\|f\|^2_{\WW^{1,2}}+\Ga(\Om,w)), \qfq i=1,\,2.$$
Thus the estimate (\ref{x4.44}) follows. \hfill$\Box$

\begin{lem}\label{l4.7}
Let the assumptions in Theorem $\ref{t4.3}$ hold. Then there is $C>0$ such that for all solutions $w\in\WW^{2,2}(\Om)$  to problem $(\ref{3.1})$
\be\Ga(\Om,w)\leq C(\|w\|^2_{\WW^{2,2}(\Om)}+\|f\|^2_{\WW^{1,2}(\Om)}).\label{4.52}\ee
\end{lem}

{\bf Proof}\,\,\,{\bf Step 1}\,\,\,We claim that for each $\varepsilon>0$ small, there is $C_\varepsilon>0$ such that
\be\sum_{j=0}^2\int_{\varepsilon}^{a-\varepsilon}|D^jw\circ\a(t,0)|^2dt\leq C_\varepsilon(\|w\|^2_{\WW^{2,2}(\Om)}+\|f\|^2_{\WW^{1,2}(\Om)}).\label{4.54}\ee

Let $t_0\in(0,a)$  be fixed and let $p_0=\a(t_0,0).$ Let $\zeta:$ $(0,\epsilon)\rw \Om$ be such that
$$\zeta(0)=p_0,\quad\zeta'(0)=-\mu(p_0),$$ where $\mu(p_0)$ is the noncharacteristic normal at the boundary point $p_0$ outside $\Om.$
From Lemma \ref{l4.3}, there are $0<\si_0<\min\{t_0,a-t_0\}$ and an an  asymptotic coordinate system  $\psi:$ $B(p_0,\si_0)\rw\R^2$
with $\psi(p_0)=(0,0)$ such that 
$$\psi(\a(t+t_0,0))=(t,-t)\qfq t\in(-\si_0,\si_0),\quad\zeta_1'(s)>0,\quad\zeta_2'(s)>0\qfq s\in(0,\varepsilon),$$ where $\psi(\zeta(s))=(\zeta_1(s),\zeta_2(s)).$
Set
$$\Om_{p_0}=\Om\cap\psi^{-1}[E(\g)],$$ where
 $$\g(t)=(t,-t),\quad E(\g)=\{\,x\,|\,-x_2<x_1<\si_1,\,\,-\si_2<x_2<\si_1\,\}.$$
  
Using (\ref{x4.24}) for the region $\Om_{p_0},$ we obtain
$$\T_1\a_t(t+t_0,0))=\pl x_1,\quad \T_2\a_t(t+t_0,0)=-\pl x_2\qfq t\in(-\si_0,\si_0),$$  
 where the operators $\T_i$ are given in $(\ref{xn4.14}).$

Observe that $w_0(x)=w\circ\psi^{-1}(x)$ is a solution to problem (\ref{3.7}) on the region $E(\g).$
Applying Proposition \ref{p3.3}, we have
\beq&&\sum_{j=0}^2\int_{-\si_1/2}^{\si_1/2}|D^jw\circ\a(t+t_0,s)|^2dt\leq C\sum_{j=0}^2\int_{-\si_1/2}^{\si_1/2}|\nabla^jw_0(t,-t)|^2dt\nonumber\\
&&\leq C(\|w_0\|^2_{\WW^{2,2}(E(\g))}+\|f\circ\psi^{-1}\|^2_{\WW^{1,2}(E(\g))})\leq C(\|w\|^2_{\WW^{2,2}(\Om)}+\|f\|^2_{\WW^{1,2}(\Om)}).\nonumber\eeq

Thus the estimates (\ref{4.54}) follows from the finitely covering theorem. By a similar argument, we have
$$\sum_{j=0}^2\int_{\varepsilon}^{b-\varepsilon}|D^jw\circ\a(t_k,s)|^2ds\leq C_\varepsilon(\|w\|^2_{\WW^{2,2}(\Om)}+\|f\|^2_{\WW^{1,2}(\Om)}),\quad k=1,\,2,$$
where $t_1=0$ and $t_2=a,$ which particularly imply that
\be \int_\varepsilon^{b-\varepsilon}|p_k'(s)|^2(b-s)ds\leq C_\varepsilon(\|w\|^2_{\WW^{2,2}(\Om)}+\|f\|^2_{\WW^{1,2}(\Om)}),\quad k=1,\,2,\label{4.55}.\ee

{\bf Step 2}\,\,\,We treat the estimates at the angular points $\a(0,0),$ $\a(0,b),$ $\a(a,0),$ and $\a(a,b),$ respectively.

Consider the angular $\a(0,b)$ first. Let $\varepsilon>0$ be given small. From Lemma \ref{l4.3}, there is an  asymptotic coordinate system  $\psi:$ $B(\a(0,b),\si_0)\rw\R^2$
with $\psi(\a(0,b))=(0,0)$ such that
$$\g(t)=\psi(\a(t,b))=(t,-t)\qfq t\in[0,\varepsilon],$$
$$\b(s)=\psi(\a(0,b-s))=(\b_1(s),\b_2(s)),\quad\b_1'(s)>0,\quad\b_2'(s)>0\qfq s\in[0,\varepsilon].$$

 Consider the region $\Om_{\a(0,b)}=\Om\cap\psi^{-1}[\Xi_1(\b,\g)].$ From (\ref{x4.24}), we have
 $$\T_2\a_s(0,b-s)=-\b_2'(s)\pl x_2\qfq s\in(0,\varepsilon).$$ 
It follows from Proposition \ref{np3.12} that
\beq\int_{b-\varepsilon_1}^b|p_1'(s)|^2(b-s)ds&&\leq C(\|f\circ\psi^{-1}\|^2_{\WW^{1,2}(\Xi_1(\b,\g))}+\|w\circ\psi_1^{-1}\|^2_{\WW^{1,2}(\Xi_1(\b,\g))})\nonumber\\
&&\leq C(\|f\|^2_{\WW^{1,2}}+\|w\|^2_{\WW^{2,2}}).\nonumber\eeq

Similarly, we can treat the estimates at the other angular points.
Thus estimate (\ref{4.52}) follows by combing the above estimates with those in Step 1.
\hfill$\Box$

{\bf Proof of Theorem \ref{t4.3}}\,\,\, Let ${\cal R}$ be the set of all $0<\omega\leq b$ such that estimate $(\ref{x4.44})$ is true. Set $\omega_0=\sup_{\omega\in{\cal R}}\omega.$ By Lemmas \ref{l4.6} and \ref{l4.7}, it is sufficient to prove
$$\omega_0\in {\cal R}.$$

By following the proof of Theorem \ref{t4.1}, we obtain a $\varepsilon>0$ small such that
$$\|w\|^2_{\WW^{2,2}(\Om(\omega_0-\varepsilon,\omega_0))}\leq C[\int_{\omega_0-\varepsilon}^{\omega_0}(|p_1'(s)|^2+|p_2'(s)|^2)(\omega_0-s)ds+\Ga(\a(\cdot,\omega_0-\varepsilon),w)+\|f\|^2_{\WW^{1,2}}],$$ where
$\Ga(\a(\cdot,\omega_0-\varepsilon), w)$ is given in (\ref{ga4.13}).
On the other hand, we fix $0<\varepsilon_1<\varepsilon$ and apply Lemma \ref{l4.7} to the region $\Om(\omega_0-\varepsilon,\omega_0-\varepsilon_1)$ to obtain
\beq\Ga(\a(\cdot,\omega_0-\varepsilon),w)&&\leq C[\|w\|^2_{\WW^{2,2}(\Om(\omega_0-\varepsilon,\omega_0-\varepsilon_1))}+\|f\|^2_{\WW^{2,2}(\Om(\omega_0-\varepsilon,\omega_0-\varepsilon_1))}]\nonumber\\
&&\leq C[\|w\|^2_{\WW^{2,2}(\Om(0,\omega_0-\varepsilon_1))}+\|f\|^2_{\WW^{2,2}(\Om(0,\omega_0-\varepsilon_1))}]\,(\mbox{ by (\ref{x4.44})})\nonumber\\
&&\leq C[\|f\|^2_{\WW^{1,2}(\Om)}+\Ga(\Om,w)].\nonumber\eeq
By Lemma \ref{l4.6}, we have $\omega_0\in{\cal R}.$ By Lemma \ref{l4.7}, we obtain $\omega_0=b.$ \hfill$\Box$

{\bf Proof of Theorem \ref{t4.4}}\,\,\,Let $w\in\Up(\Om)$ and $\varepsilon>0$ be given. We shall find a $\hat w\in\H_2(\Om)$ such that
$$\|w-\hat w\|^2_{\WW^{2,2}(\Om)}<\varepsilon.$$

Let $p_1,$ $p_2,$ and $q_0,$ $q_1$ be given in (\ref{4.3}) and (\ref{x1}), respectively.
Towards approximating $w$ by $\H(\Om)$
functions, we first approximate its traces $q_0,$ $q_1,$ $p_1,$ and $p_2.$ From Theorem \ref{t4.3}, those traces are regular except for the angular points $\a(0,0),$ $\a(0,b),$ $\a(a,0),$ and $\a(a,b).$
Next, we change their values near those angular points to make them regular and to let the $1$th order compatibility conditions hold at $\a(0,0)$ and $\a(a,0).$

{\bf Step 1}\,\,\,Consider the point $\a(0,0).$  Let $\si>0$ be given small by Lemma \ref{l4.3} such that there is an asymptotic coordinate system  $\psi:$ $B(\a(0,0),\si)\rw\R^2$
with $\psi(\a(0,0))=(0,0)$ such that
$$\psi(\a(t,0))=(t,-t)\qfq t\in[0,t_0),$$
$$\b(s)=\psi(\a(0,s))=(\b_1(s),\b_2(s)),\quad\b_1'(0)=\b_2'(0),\quad\b_1'(s)>0,\quad \b_2'(s)>0\qfq s\in[0,t_0),$$ for some  $0<t_0<\min\{a,b\}/4$ small.

From Lemma \ref{l4.3}, we have
$$p_1(s)=w_{x_2}\circ\b(s)\b_2'(s)\qfq s\in[0,t_0],$$
\be q_0'(t)=w_{x_1}(t,-t)-w_{x_2}(t,-t),\quad -\sqrt{2}q_1(t)=w_{x_1}(t,-t)+w_{x_2}(t,-t)\label{qq}\ee for $t\in(0,t_0).$
where $w(x)=w\circ\psi^{-1}(x).$ Moreover, we have
\beq D^2w(\T_1\a_t,\T_1\a_t)&&=D^2w(\pl x_1,\pl x_1)=w_{x_1x_1}(t,-t)-D_{\pl x_1}\pl x_1(w)(t,-t)=\var_{11}+\phi_{1},\nonumber\eeq and
$$D^2w(\T_2\a_t,\T_2\a_t)=\var_{22}+\phi_{2}.$$  By differential the equations in (\ref{qq}) in $t\in(0,t_0)$  and using the formulas (\ref{3.1}) and (\ref{3.5}), we obtain
$$\var_{11}=\frac{1}{2}[q_0''(t)-\sqrt{2}q_1'(t)],\quad \var_{22}=\frac{1}{2}[q_0''(t)+\sqrt{2}q_1'(t)]\qfq t\in(0,t_0),$$
$$\phi_1=\mbox{some frist order terms of $w$},\quad \phi_2=\mbox{some frist order terms of $w$}.$$
By Theorem \ref{t4.3}
$$p_1\in\WW^{1,2}(0,t_0),\quad q_0\in\WW^{1,2}(0,t_0),\quad q_1,\,\, \var_{11}t^{1/2},\,\,\var_{22},\,\,\phi_1,\,\,\phi_2\in L^2(0,t_0).$$ Thus
$$q_0''t^{1/2}=(\var_{11}+\var_{22})t^{1/2}\in L^2(0,t_0),\quad q_1't^{1/2}=\frac{1}{\sqrt{2}}(\var_{22}-\var_{11})t^{1/2}\in L^2(0,t_0).$$

We also need the following.
\begin{lem}\label{l4.8}Let
$$z(t)=\frac{1}{2}[q_0'(t)+\sqrt{2}q_1(t)]\qfq t\in(0,t_0).$$ Then $z\in \CC[0,t_0]$ and
\be p_1(0)+z(0)\b_2'(0)=0.\label{pz}\ee
\end{lem}

{\bf Proof of Lemma \ref{l4.8}}\,\,\,It follows from $z'=\var_{22}\in L^2(0,t_0)$ that $z\in\CC[0,t_0].$

We have
$$w_{x_2}\circ\b\circ\b_1^{-1}(t)-w_{x_2}(t,-t)=\int_{-t}^{\b_2\circ\b_1^{-1}(t)}w_{x_2x_2}(t,s)ds,$$ from which we obtain
$$|w_{x_2}\circ\b\circ\b_1^{-1}(t)-w_{x_2}(t,-t)|^2\leq[\b_2\circ\b_1^{-1}(t)+t]]\int_{-t}^{\b_2\circ\b_1^{-1}(t)}|w_{x_2x_2}(t,s)|^2ds.$$
For $\varepsilon>0$ given, let $\vartheta\in[\varepsilon/2,\varepsilon]$ be fixed such that
$$|w_{x_2}\circ\b\circ\b_1^{-1}(\vartheta)-w_{x_2}(\vartheta,-\vartheta)|^2=\inf_{t\in[\varepsilon/2,\varepsilon]}|w_{x_2}\circ\b\circ\b_1^{-1}(t)-w_{x_2}(t,-t)|^2.$$ Then
\beq|w_{x_2}\circ\b\circ\b_1^{-1}(\vartheta)-w_{x_2}(\vartheta,-\vartheta)|^2&&\leq\frac{2}{\varepsilon}[\b_2\circ\b_1^{-1}(\varepsilon)+
\varepsilon]\int_{\varepsilon/2}^\varepsilon\int_{-t}^{\b_2\circ\b_1^{-1}(t)}|w_{x_2x_2}(t,s)|^2ds\nonumber\\
&&\leq\si\int_{0}^\varepsilon\int_{-t}^{\b_2\circ\b_1^{-1}(t)}|w_{x_2x_2}(t,s)|^2ds\qfq t\in[\varepsilon/2,\varepsilon].\nonumber\eeq Thus, $w\in\WW^{2,2}(\Om)$ implies, by (\ref{qq}), that (\ref{pz})
holds.\\

Let $0<\varepsilon<t_0$ given small.  We shall construct $\hat q_0$ and $\hat q_1$  to satisfy the following.

(1)\,\,\,$\hat q_0(t)=q_0(t),$ $\hat q_1(t)=q_1(t)$ for $t\in[\varepsilon,a);$

(2)\,\,\,$\hat q_0\in\WW^{2,2}(0,a)$ and $\hat q_1\in\WW^{1,2}(0,a);$

(3)\,\,\,The following $1$th order compatibility conditions  hold at the point $\a(0,0),$
$$2p_1(0)+[\hat q_0'(0)+\sqrt{2}\hat q_1(0)]\b_2'(0)=0;$$

(4)\,\,\,If $\hat w\in\Up(\Om)$ is such that
$$\hat w\circ\a(t,0)=\hat q_0(t),\quad \frac{1}{2}\<D\hat w,(\T_2-\T_1)\a_t\>\circ\a(t,0)=\hat q_1(t)\qfq t\in(0,a),$$
$$ \<D\hat w,\T_1\a_s\>\circ\a(0,s)= p_1(s),\quad \<D\hat w,\T_2\a_s\>\circ\a(0,s)= p_2(s)\qfq s\in(0,b),$$
then
$$\Ga(\Om,\hat w-w)\rw0\quad\mbox{as}\quad\varepsilon\rw0.$$

For the above purposes, we define
$$\hat q_0(t)=\left\{\begin{array}{l}\si_0(\varepsilon)+[q_0'(t_0)-\int_\varepsilon^{t_0}\var_{11}ds-\int_0^{t_0}\var_{22}ds]t+\int_0^t(t-s)\var_{22}(s)ds,\quad t\in[0,\varepsilon),\\
q_0(t)\quad t\in[\varepsilon,a],\end{array}\right. $$
and
$$\hat q_1(t)=\left\{\begin{array}{l}q_1(t_0)+\frac{1}{\sqrt{2}}\int_\varepsilon^{t_0}\var_{11}ds-\frac{1}{\sqrt{2}}\int_t^{t_0}\var_{22}ds,\quad t\in(0,\varepsilon),\\
q_1(t),\quad t\in[\varepsilon,a],\end{array}\right.$$ where
$$\si_0(\varepsilon)=q_0(\varepsilon)-q_0'(\varepsilon)\varepsilon+\int_0^\varepsilon s\var_{22}(s)ds.$$

Clearly, (1) and (2) hold for the above $\hat q_0$ and $\hat q_1.$
 Since
$$q_0'(t_0)-\int_\varepsilon^{t_0}\var_{11}ds-\int_0^{t_0}\var_{22}ds=q_0'(\varepsilon)-\int_0^\varepsilon\var_{22}(s)ds=q_0'(\varepsilon)-z(\varepsilon)+z(0),\qfq t\in(0,\varepsilon),$$
$$\frac{1}{\sqrt{2}}\int_\varepsilon^{t_0}\var_{11}ds-\frac{1}{\sqrt{2}}\int_t^{t_0}\var_{22}ds=q_1(\varepsilon)-q_1(t_0)+\frac{1}{\sqrt{2}}[z(t)-z(\varepsilon)],$$
using (\ref{pz}), we have
$$2p_1(0)+[\hat q_0'(0)+\sqrt{2}\hat q_1(0)]\b_2'(0)=q_0'(\varepsilon)+\sqrt{2}q_1(\varepsilon)-2z(\varepsilon)=0.$$
Next, we check (4). It follows that
\beq|q_0(t)-\hat q_0(t)|^2&&=|\int_t^\varepsilon\int_s^\varepsilon q_0''(\tau)d\tau ds+\int_t^\varepsilon(t-s)\var_{22}(s)ds|^2\nonumber\\
&&\leq2(\varepsilon-t+t\ln\frac{t}{\varepsilon})\int_0^\varepsilon|q_0''(\tau)|^2\tau d\tau+\frac{2}{3}\varepsilon^3\int_0^\varepsilon|\var_{22}(s)|^2ds\qfq t\in(0,\varepsilon). \nonumber\eeq
In addition,
$$|q_0'(t)-\hat q_0'(t)|^2=|\int_t^\varepsilon\var_{11}(s)ds|^2\leq(\ln\frac{\varepsilon}{t})\int_0^\varepsilon|\var_{11}(s)|^2sds\qfq t\in(0,\varepsilon).$$ Similarly, we have
$$|q_1(t)-\hat q_1(t)|^2\leq(\ln\frac{\varepsilon}{t})\int_0^\varepsilon|\var_{11}(s)|^2sds\qfq t\in(0,\varepsilon).$$

Using (\ref{ga4.13}) and the above estimates, we have
\beq\Ga(\a(\cdot,0),w-\hat w)&&=\sum_{j=0}^1\|D(w-\hat w)\circ\a(\cdot,0)\|^2_{L^2(0,\varepsilon)}+\int_0^\varepsilon[|D^2(w-\hat w)(\T_1\a_t,\T_1\a_t)|^2t\nonumber\\
&&\quad+|D^2(w-\hat w)(\T_2\a_t,\T_2\a_t)|^2(a-t)]dt\nonumber\\
&&\leq C\int_0^\varepsilon(|q_0(s)-\hat q_0(s)|^2+|q_0'(t)-\hat q_0'(t)|^2+|q_1(t)-\hat q_1(t)|^2\nonumber\\
&&\quad+|\var_{11}-\hat\var_{11}|^2t+|\var_{22}-\hat\var_{22}|^2)dt\nonumber\\
&&\leq C\int_0^\varepsilon[(|q_0''(t)|^2+|\var_{11}(t)|^2)t+|\var_{22}(t)|^2]dt,\eeq where
$$\hat\var_{11}=\frac{1}{2}[\hat q_0''(t)-\sqrt{2}\hat q_1'(t)]=0,\quad \hat\var_{22}=\frac{1}{2}[\hat q_0''(t)+\sqrt{2}\hat q_1'(t)]=\var_{22}.$$
Thus (4) follows.

{\bf Step 2}\,\,\,As in Step 1, we change the values of $q_0$ and $q_1$ near the point $\a(a,0)$ to get $\hat q_0$ and $\hat q_1$ in $\WW^{2,2}(0,a)$ and in $\WW^{1,2}(0,a),$ respectively, such that the $1$th order compatibility conditions at $\a(a,0)$ hold to approximate $q_0$ and $q_1.$ Then we change the values of $p_1$ and $p_2$ near the points $\a(0,b)$ and $\a(a,b),$ respectively, such that $\hat p_1,$ $\hat p_2\in\WW^{1,2}(0,b)$ approximate $p_1$ and $p_2,$ respectively.
Thus the proof completes from Theorem \ref{t4.2}.  \hfill$\Box$

\setcounter{equation}{0}
\section{Proofs of Main Results in Section 1}\label{s5}
\def\theequation{5.\arabic{equation}}
\hskip\parindent
{\bf Proof of Theorem \ref{t00}}\,\,\, Let $\Om\subset M$ be a noncharacterisic  region of class $\CC^{2,1}.$
For $U\in \CC^{1,1}(\Om,T^2_{\sym})$ given, we consider problem
\be\sym\nabla y=U\quad\mbox{on}\quad \Om.\label{5n.1}\ee

(1)\,\,\,Consider problem
\be\<D^2v,Q^*\Pi\>=P(U)-2v\kappa\tr_g\Pi+X(v)\qfq x\in\Om,\label{5.5}\ee where $P(U)$ and $X$ are given in (\ref{2.19}) and (\ref{2.20}), respectively, with
the boundary data
\be\<Dv,\T_2\a_s\>\circ\a(0,s)=\<Dv,\T_2\a_s\>\circ\a(a,s)=0\qfq s\in(0,b),\label{nn5.5}\ee
\be v\circ\a(t,0)=\frac{1}{\sqrt{2}}\<Dv,(\T_2-\T_1)\a_t\>\circ\a(t,0)=0\qfq t\in(0,a),\label{nn5.6}\ee where $\T_1$ and $\T_2$ are given in (\ref{xn4.14}).

Since
$$P(U)\in L^\infty(\Om),\quad X\in L^\infty(\Om),$$ it follows from Theorem \ref{t4.1} that problem (\ref{5.5}) with the data (\ref{nn5.5}) and (\ref{nn5.6}) has a unique solution
$v\in\CC^{0,1}(\Om)$ with the bounds
\be \|v\|_{\CC^{0,1}(\Om)}\leq C\|U\|_{\CC^{1,1}(\Om,T^2_{\sym})}.\label{5.7}\ee
From Theorem \ref{t2.1}, there is a solution $y\in\CC^{0,1}(\Om,\R^3)$ to (\ref{5n.1}). Let
$$w=\<y,\n\>,\quad W=y-w\n.$$
Then $w\in\CC^{0,1}(\Om).$ It follows from \cite[lemma 4.3]{Yao 2011} that $W\in\CC^{1,1}(\Om,T)$ and (\ref{t0}) holds.

(2)\,\,\,Let $\Om\in\CC^{m+2,1}$ and $U\in\CC^{m+1,1}(\Om,T^2_{\sym})$  be given for some $m\geq1.$ Let
$$q_0(t)=q_1(t)=0\qfq t\in[0,a].$$ Let $\Q_k\Big(0,0,P(U)\Big)(t)$ be given in the formula (\ref{xn4.7}) for $t\in[0,a]$ and $1\leq k\leq m-1.$ We define
\be \phi_j(s)=\left\{\begin{array}{l} 0,\quad m=1,\\
\sum_{l=1}^{m-1}\frac{p_j^{(l)}(t_j)}{l!}s^l,\quad m\geq2,\end{array}\right.\qfq s\in[0,b],\quad j=1,\,\,2,\label{5.8}\ee where $p_j^{(l)}(t_j)$ are given by the right hand sides of (\ref{xn4.9}) for $1\leq l\leq m-1$ and $1\leq j\leq2,$ where $q_0=q_1=0$ and $f=P(U).$
Clearly, the $m$th
compatibility conditions hold true for the above $q_0,$ $q_1,$ $\phi_1,$ $\phi_2,$ and $P(U).$
From Theorem \ref{t4.1}, there is a solution  $v\in \CC^{m,1}(\overline\Om)$ to problem $(\ref{5.5})$ with the data
$$\<Dv,\T_2\a_s\>\circ\a(0,s)=\phi_1(s),\quad \<Dv,\T_2\a_s\>\circ\a(a,s)=\phi_2(s)\qfq s\in(0,b),$$
$$ v\circ\a(t,0)=\frac{1}{\sqrt{2}}\<Dv,(\T_2-\T_1)\a_t\>\circ\a(t,0)=0\qfq t\in(0,a).$$ Moreover, it follows from (\ref{xn4.10}) and (\ref{2.19}) that
$$ \|v\|_{\CC^{m,1}(\overline{\Om})}\leq C\|U\|_{\CC^{m+1,1}(\Om,T^2)},$$ which implies the estimate (\ref{t2}) is true.  \hfill$\Box$

{\bf Proof of Theorem \ref{t1.1}}\,\,\,
Let
$$V=W+w\n,\quad w=\<V,\n\>.$$ The regularity of
$$\sym DW=-w\Pi\in\WW^{2,2}(\Om,\R^3)$$ implies
$$W\in\WW^{3,2}(\Om,T).$$

Let $E_1,$ $E_2$ be a frame field on $\Om$ with the positive orientation and let
$$v=\frac{1}{2}[\nabla V(E_2,E_1)-\nabla V(E_1,E_2)].$$ From Theorem \ref{t2.1} $v$ is a solution to problem
\be\<D^2v,Q^*\Pi\>=-2v\kappa\tr_g\Pi+X(v)\qfq x\in\Om,\label{5.1}\ee where $\kappa\tr_g\Pi\in\CC^{m,1}(\overline{\Om})$ and
$X=(\nabla\n)^{-1}D\kappa\in\CC^{m-1,1}(\overline{\Om},T),$ where $\CC^{-1,1}(\overline{\Om},T)=L^\infty(\Om,T).$

It is easy to check that
$$\nabla_{E_i}V=D_{E_i}W+w\nabla_{E_i}\n+[E_i(w)-\Pi(W,E_i)]\n\qfq i=1,\,\,2.$$ Thus
$$v=DW(E_2,E_1)-DW(E_1,E_2)\in\WW^{2,2}(\Om).$$
From Theorems \ref{t4.4}, \ref{t4.1}, and \ref{t4.2}, there are solutions $v_n\in\CC^{m,1}(\overline{\Om})$ to problem (\ref{5.1}) such that
$$\lim_{n\rw\infty}\|v_n-v\|_{\WW^{2,2}(\Om)}=0.$$
Let
$$u_n=-Q(\na\n)^{-1}Dv_n,\quad u=-Q(\na\n)^{-1}Dv.$$ Then $u_n\in\CC^{m-1,1}(\overline{\Om}).$

From Theorem \ref{t2.1} (see (\ref{2.7})), there exist $\hat V_n\in\CC^{m,1}(\overline{\Om},\R^3)$  such that
\be\left\{\begin{array}{l}\nabla_{E_1}\hat V_n=v_nE_2+\<u_n,E_1\>\n,\\
\nabla_{E_2}\hat V_n=-v_nE_1+\<u_n,E_2\>\n,\end{array}\right.\qfq n=1,\,\,2,\,\,\cdots.\nonumber\ee

Define
$$V_n(\a(t,s))=\hat V_n(\a(0,s))-\hat V_n(\a(0,0))+V(\a(0,0))+\int_0^t\nabla_{\a_t}\hat V_ndt\qfq n=1,\,\,2,\,\,\cdots.$$ Thus $V_n\in\V(\Om,\R^3)\cap\CC^{m,1}(\Om,\R^3))$  satisfy (\ref{sl}). \hfill$\Box$

{\bf Proof of Theorem \ref{t1.2}}\,\,\,As in $\cite{HoLePa}$ we conduct in $2\leq i\leq m.$  Let
$$u_\varepsilon=\sum_{j=0}^{i-1}\varepsilon^jw_j$$ be an $(i-1)$th order isometry of class $\CC^{2(m-i+1)+1,1}(\overline{\Om},\R^3),$ where $w_0=\id$ and $w_1=V$ for some $i\geq2.$ Then
$$\sum_{j=0}^k\na^Tw_j\na w_{k-j}=0\qfq 0\leq k\leq i-1.$$

Next, we shall find out $w_i\in\CC^{2(m-i)+1,1}(\overline{\Om},\R^3)$ such that
$$\phi_\varepsilon=u_\varepsilon+\varepsilon^iw_i$$ is an $i$th order isometry.
From Theorem $\ref{t00}$ there exists  a solution $w_i\in\CC^{2(m-i)+1,1}(\overline{\Om},\R^3)$ to problem
$$\sym \nabla w_i=-\frac{1}{2}\sym\sum_{j=1}^{i-1}\sym\na^Tw_j\na w_{i-j}$$ which satisfies
\beq\|w_i\|_{\CC^{2(m-i)+1,1}(\overline{\Om},\R^3)}&&\leq C\|\sum_{j=1}^{i-1}\sym\na^Tw_j\na w_{i-j}\|_{\CC^{2(m-i+1),1}(\overline{\Om},\R^3)}\nonumber\\
&&\leq C\sum_{j=1}^{i-1}\|w_j\|_{\CC^{2(m-i+1)+1,1}(\overline{\Om},\R^3)}\|w_{i-j}\|_{\CC^{2(m-i+1)+1,1}(\overline{\Om},\R^3)}.\nonumber\eeq
The conduction completes.\hfill$\Box$

{\bf Theorem \ref{t1.3}} will follow from the density of the Sobolev space and Proposition \ref{p5.1} below.

\begin{pro}\label{p5.1}Let $\Om\subset M$ be a noncharacteristic region of class $\CC^{2,1}.$ Then for $U\in\WW^{3,2}(\Om,T^2_{\sym})$ there exits a solution $w\in\WW^{2,2}(\Om,\R^3)$ to problem
$$\sym\nabla w=U.$$
\end{pro}

{\bf Proof}\,\,\,Consider problem (\ref{5.5}) with the data (\ref{nn5.5}) and (\ref{nn5.6}). By (\ref{cn}) the first order compatibility conditions hold. Since $P(U)\in \WW^{1,2}(\Om),$ the proposition follows from Theorems \ref{t4.2} and \ref{t2.1}. \hfill$\Box$

{\bf Proof of Theorem \ref{t1.6}}\,\,\,A recovery sequence can be constructed, based on Theorems \ref{t1.1} and \ref{t1.2}, as in  the proof of \cite[Theorem 6.2]{HoLePa}.
We present a skeleton of the proof. For the further details, see \cite{HoLePa}.

From the density of Theorem \ref{t1.1} and  the continuity of the functional $I$ with
respect to the strong topology of $\WW^{2,2},$ we can assume $V\in\V(\Om,\R^3)\cap\CC^{2m-1,1}(\Om,\R^3).$

{\bf Step 1}\,\,\,Let $\varepsilon=\frac{\sqrt{e^h}}h$ so $\varepsilon\rw0,$ as $h\rw0,$ by assumption (\ref{x1.3}). Therefore, by Theorem \ref{t1.2}, there exists a sequence
$w_\varepsilon:$ $\overline{\Om}\rw\R^3,$ equibounded in $\CC^{1,1}(\Om,\R^3),$ for all $h>0,$
$$u_\varepsilon=\id+\varepsilon V+\varepsilon^2w_\varepsilon$$ is a $m$th isometry of class $\CC^{1,1}.$ Then
$$\varepsilon^{m+1}=\o(\sqrt{e^h}).$$
Consider the sequence of deformations $u_h\in\WW^{1,2}(\Om_h,\R^3)$ defined by
$$u_h(x+t\n)=u_\varepsilon(x)+t\n_\varepsilon(x)+\frac{t^2}2\varepsilon d_h(x)\qfq x+t\n(x)\in\Om_h,$$ where $\n_\varepsilon(x)$ denotes the unit normal to $u_\varepsilon(\Om)$ at $u_\varepsilon(x)$
and $d_h\in\WW^{1,\infty}(\Om,\R^3)$ is such that $\lim_{h\rw0}h^{1/2}\|d_h\|_{\WW^{1,\infty}}=0$ and
$$\lim_{h\rw0}d_h(x)=2c(x,\sym(\nabla(A\n)-A\Pi)_{\tan})\qfq x\in\Om,$$ where $c(x,F_{\tan})$ denotes the unique vector satisfying $\Q_2(x,F_{\tan})=\Q_3(F_{\tan}+c\otimes\n(x)+\n(x)\otimes c).$ We have
$$\n_\varepsilon(x)=\n(x)+\varepsilon A\n+\O(\varepsilon^2).$$

{\bf Step 2}\,\,\,We have
$$\frac{E_h(u_h)}{e^h}=\frac{1}{e^hh_0}\int_{-h_0/2}^{h_0/2}\int_\Om W(\nabla_hy^h(x+t\n(x)))(1+\frac{th}{h_0}\tr_g\Pi+\frac{t^2h^2}{h_0^2}\kappa)dgdt,$$
where $\nabla_hy^h(x+t\n(x))=\nabla u_h(x+\frac{th}{h_0}\n).$ Let
$$K^h(x+t\n(x))=(\nabla_hy^h)^T\nabla_hy^h-\Id.$$ Using the formulas $\nabla^Tu_\varepsilon\nabla u_\varepsilon=\Id+\O(\varepsilon^{m+1})=\Id+\o(\sqrt{e^h})$ and $h\varepsilon=\sqrt{e^h},$ we have
$$K^h_{\tan}=2\frac{t\sqrt{e^h}}{h_0}(\Id+\frac{th}{h_0}\Pi)^{-1}\sym (\nabla(A\n)-A\Pi)(\Id+\frac{th}{h_0}\Pi)^{-1}+\o(\sqrt{e^h}),$$
$$\<K^h\n,\n\>=2\frac{t\sqrt{e^h}}{h_0}\<\n_\varepsilon,d_h\>+\o(\sqrt{e^h}),$$
$$\<K^h\a,\n\>=\frac{t\sqrt{e^h}}{h_0}\<\nabla u_\varepsilon(\Id+\frac{th}{h_0}\Pi)^{-1}\a,d_h\>+\o(\sqrt{e^h})\qfq \a\in T_x\Om.$$
Then
$$\lim_{h\rw0}\frac{K^h_{\tan}}{2\sqrt{e^h}}=\frac{t}{h_0}\sym (\nabla(A\n)-A\Pi)\qiq L^\infty(\Om_{h_0}),$$
$$\lim_{h\rw0}\frac{K^h\n}{2\sqrt{e^h}}=\frac{2t}{h_0}c(x,\sym(\nabla(A\n)-A\Pi)_{\tan})\qiq L^\infty(\Om_{h_0}).$$

{\bf Step 3}\,\,\,We have
$$\frac{W(\nabla_yy_h)}{e^h}=\frac12\Q_3(\frac{K^h}{2\sqrt{e^h}}+\frac{1}{\sqrt{e^h}}\O(|K^h|^2)+\frac{1}{e^h}\o(|K^h|^2).$$
Then the limit (\ref{lim}) follows from Step 2.
\hfill$\Box$

{\bf Compliance with Ethical Standards}

Conflict of Interest: The author declares that there is no conflict of interest.

Ethical approval: This article does not contain any studies with human participants or animals performed by the author.

 \end{document}